\documentstyle[11pt,aaspp4]{article}

\lefthead{Marzke et al.}
\righthead{Luminosity Function at $z \le 0.05$}

\begin{document}
\title{The Galaxy Luminosity Function at $z \le 0.05$: Dependence on Morphology}
\def\bu{$\bullet$ \hskip 7pt}
\def\listitem{\par \hangindent=50pt\hangafter=1
     $\ $\hbox to 20pt{\hfil $\bullet$ \hfil}}
\def\refitem{\par\noindent\hangindent 20pt}
\def\ref2{\par\noindent\hangindent 40pt}
\def\mb{$m_{B(0)}$}
\def\MB{$M_{B(0)}$}
\def\puncspace{\ifmmode\,\else{\ifcat.\C{\if.\C\else\if,\C\else\if?\C\else%
\if:\C\else\if;\C\else\if-\C\else\if)\C\else\if/\C\else\if]\C\else\if'\C%
\else\space\fi\fi\fi\fi\fi\fi\fi\fi\fi\fi}%
\else\if\empty\C\else\if\space\C\else\space\fi\fi\fi}\fi}
\def\SP{\let\\=\empty\futurelet\C\puncspace}
\def\iras{{\it IRAS}\SP}
\def\dec{$\delta$\SP}
\def\ra{$\alpha$\SP}
\def\menor {$\leq$\SP}
\def\maior {$\geq$\SP}
\def\kms{\rm ~km~s^{-1}}
\def\BT{B$_{T}$\SP}
\def\degree{$^\circ$\SP}
\def\h1{$h^{-1}$\SP}
\def\hour{$^h$\SP}
\def\arcm{'\SP}
\def\ap{$\approx$\SP}
\def\minutes{$^m$\SP}
\def\etal{{\it et al.\/}\ }
\def\ie{\it i.e.\/\rm,\ }
\def\eg{\it e.g.\/\rm,\ }
\def\apj{ApJ,}
\def\apjlet{Ap. J. (Letters)}
\def\apjsup{ApJS,}
\def\aa{Astr. Ap.}
\def\mnras{MNRAS,}
\def\araa{Ann. Rev. Astr. Ap.}
\def\aj{AJ,}

\def\hmpc{h^{-1}\,{\rm Mpc}}
\def\b26{m_{SSRS2}}

\author{Ronald O. Marzke\altaffilmark{1}} 
\affil{
Carnegie Observatories, 813 Santa Barbara St., Pasadena, CA 91101
\\ and \\
Dominion Astrophysical Observatory, Herzberg Institute of Astrophysics,
National Research Council of Canada, 5071 W. Saanich Rd., Victoria, BC, Canada
V8X 4M6 
}
\author{L. Nicolaci da Costa\altaffilmark{2}}
\affil{European Southern Observatory, Karl-Schwarzchild-Strasse 2, 85748
Garching bei Munchen}
\author{Paulo S. Pellegrini and Christopher N.A. Willmer}
\affil{Departamento de Astronomia CNPq/Observatorio Nacional, Rua General Jose
Cristino, 77, Rio de Janeiro, Brazil 20.921}
\author{Margaret J. Geller}
\affil{Harvard-Smithsonian Center for Astrophysics, 60 Garden St.,
Cambridge, MA 02138}


\altaffiltext{1}{Hubble Fellow}

\altaffiltext{2}{Also Departamento de Astronomia CNPq/Observatorio Nacional, Brazil}


\begin{abstract}

We investigate the dependence of the local galaxy luminosity
function on morphology using 5,404 galaxies from the recently enlarged
Second Southern Sky Redshift Survey (SSRS2).
Over the range $-22 \le M_{B} \le -14$ ($H_0 = 100\,\,\rm km\, s^{-1}\,Mpc^{-1}$), 
the luminosity function of early-type galaxies
is well fitted by a Schechter function with parameters
$M_*=-19.37^{+0.10}_{-0.11}$, $\alpha=-1.00^{+0.09}_{-0.09}$ 
and $\phi_*=4.4\pm 0.8 \times 10^{-3} \,
\rm{Mpc}^{-3}$.  The spiral luminosity function is very similar and is well fitted by the
parameters $M_*=-19.43^{+0.08}_{-0.08}$, $\alpha=-1.11^{+0.07}_{-0.06}$ 
and $\phi_*=8.0 \pm 1.4 \times 10^{-3} \, \rm{Mpc}^{-3}$ over the same range in absolute magnitude.

The flat faint end of the early-type luminosity function is consistent with
earlier measurements from the CfA Redshift Survey (Marzke \etal 1994a) but
is significantly steeper than the slope measured in the Stromlo-APM survey 
(Loveday \etal 1992).  Combined with the increased normalization of the overall
LF measured from intermediate redshift surveys, the flat faint-end slope of the 
E/S0 LF produces no-evolution models which reproduce the deep HST counts of early-type galaxies
remarkably well.  However, the observed normalization of the SSRS2 LF is consistent with the low value measured
in other local redshift surveys. The cause of this low-redshift anomaly
remains unknown.

The luminosity function of irregular and peculiar galaxies in the SSRS2 is 
very steep: 
$M_* =-19.78^{+0.40}_{-0.50}$, $\alpha=-1.81^{+0.24}_{-0.24}$ and $\phi_*=0.2 \pm 0.08 \times 10^{-3} \,
\rm{Mpc}^{-3}$.
The steep slope at the faint end is consistent with the LFs measured for Sm-Im galaxies in the CfA survey,
UV-selected galaxies (Treyer \etal 1997), star-forming field galaxies (Bromley \etal 1997)
and the bluest galaxies in the SSRS2 (Marzke \& da Costa 1997).
As shown by Driver, Windhorst \& Griffiths (1995$b$), the steep LF 
reduces the observed excess of faint irregulars over no-evolution predictions
but cannot explain it entirely.
 
\end{abstract}




%

\section{Introduction}

The luminosity function (LF) of low-redshift galaxies is a primary
benchmark for models of galaxy formation.
The detailed shape of the LF reflects a host of physical
processes ranging from the collapse of dark-matter
halos to the complex cycle of gas cooling, star
formation, and feedback into the interstellar
medium.  The dependence of the LF on other observables such as morphology, color
and emission-line strength provide further constraints on the input physics.
Along with dynamical measures of galaxy
masses, the LF for various types along the Hubble sequence
anchors the theory of galaxy formation at low redshift.

Deep counts and redshift surveys of faint galaxies provide
a wealth of constraints on galaxy evolution.
Although recent surveys have been designed to cover a wide
range in redshift and apparent magnitude
(\eg Lilly \etal 1995 and Ellis \etal 1996),
limits on available telescope time force 
a compromise between the redshift baseline and
the sampling at each redshift.  Surveys targeting intermediate-redshift galaxies
are inevitably quite sparse at redshifts less than $z \approx 0.1$.
The Autofib survey (Ellis \etal 1996) is perhaps the best compromise;
the solid angle of that survey decreases with limiting magnitude
in order to produce a catalog with similar numbers of galaxies 
in each broad redshift bin.
However, in the regime $z \le 0.1$, wide-angle surveys at
relatively bright apparent magnitudes remain the best approach
for measuring the local LF and its variations from one part
of the local universe to another.

Very local samples ($z \le 0.05$) allow the most detailed investigations
of individual galaxy properties at the present-epoch; nearby galaxies are conveniently large and bright.
The cost of this acuity is the unavoidable fact that smaller volumes
are less likely to be fair samplings of the universe.
Because fluctuations in the density
field decrease with scale, samples covering larger volumes 
yield better estimates of the galaxy density.  Furthermore, 
the properties of individual galaxies 
correlate with local density on relatively small scales (Dressler
1980, Postman \& Geller 1984, Maia \& da Costa 1990, Hashimoto \etal 1997).
Correlations between galaxy properties and density on large scales
are only vaguely understood (Park \etal 1994, Willmer, da Costa \& Pellegrini 1997), but it
seems clear that fair estimates of the local LF and its dependence
on galaxy properties requires averaging over a large number of
large-scale structures.  Limits to our knowledge of the present-epoch
universe are fundamental: in order to resolve the evolutionary history
of the galaxy population from $\sim 1\,\rm Gyr$ ago up to the present
day, we are consigned
by light-travel time alone to a region $\sim 600 \hmpc$ across.  

Recent HST surveys of faint field galaxies have opened new windows on galaxy
morphology at high redshift (Driver \etal 1995$a,b$, Glazebrook \etal 1995, Abraham \etal 1996).  
Because different types of galaxies have
quite different spectral energy distributions, apparent magnitude-limits
impose redshift-dependent filters
upon the observed distribution of galaxy types.  Predictions of the
morphological composition of intermediate and high-redshift samples require (at the
very least) a detailed understanding of the distribution of morphologies nearby.

In this paper, we measure the luminosity function 
and its dependence on galaxy morphology using
the 5404 galaxies in the combined SSRS2 samples
in the northern and southern Galactic caps.
We summarize the data in $\S2$
and review our computational techniques in $\S3$.
Section $\S4$ summarizes the results, and in
$\S5$ we discuss the implications for deep
HST counts.  We conclude in $\S6$.

\section{Data}

The details of the SSRS2 southern sample have appeared in other papers (Alonso \etal 
1993a,b, da Costa \etal 1994, Marzke \& da Costa 1997).  The sample has since
been enlarged to include 1,937 galaxies from the northern Galactic
hemisphere.  A comprehensive description
of the full sample may be found along with the entire SSRS2 catalog in da Costa \etal (1998).
The photometric sample is generated from
the non-stellar sources of the STScI Guide Star Catalog
(GSC hereafter, Lasker \etal 1990).  Galaxies were distinguished from stars, HII regions and other
contaminants first by matching with existing galaxy catalogs and then by 
careful examination of each unmatched source by eye.  The algorithm for determining
local sky background in the GSC imposed a maximum size for galaxy detection.  This size
was appoximately 10$'$.  Galaxies larger than this make up a very small fraction of the 
magnitude-limited sample, but for completeness, they were inserted into the catalog by hand
using the Morphological Catalog of Galaxies (Vorontsov-Velyminov \& Karanchentsev 1963-1969) 
and the ESO Surface Photometry Catalog (Lauberts \& Valentijn 1989).  It should be noted that
the exclusion of very large galaxies is endemic to large plate surveys such as the APM (Maddox \etal 1990)
and COSMOS (Heydon-Dumbleton \etal 1989).  
The magnitude system is calibrated with CCD photometry in Alonso \etal 1993a,b, where the
magnitudes $b_{SSRS2}$ are defined to match the B(0) system used in the CfA
Survey as closely as possible.  In practice, $b_{SSRS2}$ turns out to be very close
to the flux within the 26 mag arcsec$^{-2}$ isophote, or $B_{26}$
on the ESO-LV system.  The full sample now includes redshifts for all 5,426 
galaxies brighter than $\b26=15.5$ over 1.69 steradians of the southern sky.  
The boundaries of the survey are defined as follows: $-40^\circ \le \delta \le -2.5^\circ$
and $b_{II} \le -40^\circ$ for SSRS2 South, $\delta \le 0^\circ$
and $b_{II} \ge 35^\circ$ for SSRS2 North.  Because the sampling beyond $z=0.05$
becomes somewhat sparse, we restrict our computations of the luminosity
function to $z \le 0.05$.

Morphological classifications in the SSRS2 come from several sources.
The accuracy of the types varies from the detailed morphologies of Corwin \etal (1985)
to rough designations assigned by one of the authors (PSP) using film copies of the
ESO $B$ plates (and in some cases with further examination of SERC J copies).  
The various sources and the modifications required to establish a homogeneous
system are described in detail in da Costa \etal (1997).
The Corwin morphologies are accurate to approximately one
de Vaucouleurs $T$ type, while the roughest classifications distinguish only the principal types:
E, S0, spiral, and irregular.  For the purposes of this paper, we smooth all the classifications
into three broad categories in order to assure a consistent morphological scale:
E/S0, spiral, and irregular/peculiar/interacting.  Only 22 of the 5426 SSRS2 galaxies
could not be classified on this scale.  In order to 
give some idea of the accuracy of our morphologies, we show a representative range of galaxies
at each morphology in Figure 1.  The three columns contain (from left to right) the best, 
typical, and worst examples of each type of galaxy in the SSRS2.  The ``best" and ``worst"
candidates were chosen to represent roughly the best and worst 10\% of the sample.
It is clear that in the ``difficult" bin, the finer classifications blur; for example,
faint irregulars may be labeled peculiar and {\it vice versa}.  However, even for these
very faint galaxies, the distinction between E/S0, spiral and irregular/peculiar/merger
is still relatively straightforward.  We discuss the possible effects of classification
errors in \S 5.

\section{Technique}

We compute the luminosity function for different morphological types
using the maximum likelihood techniques of Sandage, Tammann \& Yahil (1979, hereafter STY) and 
Efstathiou, Ellis \& Peterson (1988, hereafter EEP).  In the STY approach, we fit
a Schechter function to each luminosity function
$$
\phi (M) = 0.4\ln 10 \, \phi_* {\lbrack 10^{0.4(M_*-M)} \rbrack} ^{1+\alpha}
\,{\rm exp} {\lbrack 10^{0.4(M_*-M)} \rbrack }
$$
by maximizing the likelihood $\cal L$ 
that all sample galaxies appear in a magnitude-limited
redshift survey,
$$
{\cal L} = {\prod_{i=1}^N} {\phi (M_i) \over {\int_{-\infty}^{M_{min}(z_i)} \phi (M) \,dM}}
$$
Here, each $M_i$ is a measured absolute magnitude (corrected for Galactic extinction 
and the K-correction at $z_i$), and $M_{min}(z_i)$ is the
faintest observable absolute magnitude given the redshift of galaxy $i$, the $K$-correction
at that redshift, the Galactic extinction and
the apparent magnitude limit of the sample.  The type-dependent $K$-corrections 
come from the model spectral energy distributions of Rocca-Volmerange \& Guiderdoni
(1988), and extinction corrections are $4.0 E(B-V)$ using the reddening measurements of Burstein
\& Heiles (1982).  We compute all distances using $H_0=100 \,\,\rm km \,s^{-1}\,Mpc^{-1}$ and $q_0=0.2$.  
At these low redshifts, $q_0$ has little effect on the computed absolute magnitudes; our choice is meant to 
reflect recent observational constraints on the value of the mean mass density and does
not include any contribution from a cosmological constant.
Because of possible distance errors,
we ignore all galaxies with Hubble velocities less than 500 km s$^{-1}$ after
correction to the local group barycenter.  We investigate residual effects of local
peculiar velocities in \S3.

Because the normalization of the LF drops out of the likelihood function, the STY 
technique yields unbiased estimates of the 
shape parameters $M_*$ and $\alpha$ even in the presence of large
density inhomogeneities (as long as the Schechter function is a
reasonable match to the shape of the LF and the correlations
between the LF and the density field are weak).  We compute confidence
intervals on the shape parameters by computing the locus of points
in the $M_*$-$\alpha$ plane where
$ \ln {\cal L}_\beta = \ln {\cal L}_{max} - \chi_\beta$, where $\chi_\beta$
is the beta point of the $\chi^2$ distribution (EEP).  Finally,
we compute the normalization $\phi_*$ using the minimum-variance
estimate of the mean density in redshift shells (Davis \& Huchra 1982).  Because there
are large density fluctuations on the scale of the SSRS2, we simply compute
the mean value of $\phi_*$ between 3,000 and 12,000 km s$^{-1}$ in bins of 500 km s$^{-1}$.
We compute the uncertainty in the mean by combining the standard deviation
of density estimates in the redshift histogram with the uncertainty in
the selection function. 

The stepwise maximum-likelihood method of EEP approximates the LF as a 
set of step functions.  The values of the steps at each absolute magnitude
form the set of fitted parameters 
(EEP, Loveday \etal 1992, Marzke \etal 1994$a$,Marzke, Huchra \& Geller 1994$b$,
Lin \etal 1996, Heyl \etal 1997).  Once again, the shape of the LF is determined independently
of the density field.  In this approach, error bars on each step come from
the diagonal elements of the inverted information matrix, which consists
of second derivatives of the likelihood function.

\section{The Luminosity Functions}

Figure 2 shows LFs for the three broad galaxy classes discussed in \S2: E/S0, Spiral and
Irregular/Peculiar.
The dashed line also gives the LF of the combined sample.  Confidence intervals for the Schechter
shape parameters appear in Figure 3, and the first four rows of Table 1 list the
fitted parameters along with $1\sigma$ errors.  The fitted SWML parameters are listed
in Table 2 along with the number of galaxies in each bin.  Note that the redshift and
absolute magnitude limits reduce the total number of galaxies used in the LF computation
to 5,036.  Two conclusions are
immediately evident from Figures 2 and 3.  First, the faint end of the
E/S0 LF is flat ($\alpha=-1.00 \pm 0.09$)
to the absolute magnitude limit of our survey, $M_B=-14$.  Second, the
irregular/peculiar LF is steep: $\alpha=-1.81 \pm 0.24$.  As we will discuss in
the next section, both of these conclusions are relevant to the interpretation 
of deep galaxy counts.  Before we proceed, however, it is worth looking more
carefully at some of the systematic uncertainties affecting these luminosity functions.

The faint end of the luminosity function of any type of galaxy is
notoriously difficult to measure.  In a magnitude-limited redshift survey,
intrinsically faint galaxies represent a tiny fraction
of the final sample.  The faintest galaxies
can only be detected in a relatively small, nearby volume
which is subject to large density fluctuations.  
A further complication
in the analysis of very nearby galaxies is the distance
uncertainty introduced by peculiar velocities.  Although
uncertainty in the Hubble constant alone does not affect
the $shape$ of the luminosity function, systematic deviations
from the Hubble flow certainly can.

The redshift maps in Figure 4 demonstrate some of these difficulties.
In this figure, right ascension is the angular coordinate, and the
radial coordinate represents redshift.  The left column shows the entire
redshift range; the center of the circle is $cz=0$ and the outer boundary
is $cz=20,000 \kms$.  Figure 4$a$ shows the declination range $-40^\circ \le
\delta \le -20^\circ$ while Figure 4$c$ shows the range $-20^\circ \le
\delta \le 0^\circ$.  We include these plots to give an idea of both the
characteristic structures and the boundaries of the survey (because of the 
broad slices in declination, the Galactic
latitude cut is difficult to show directly).
The right-hand column is an expanded view of the region $cz \le 3,000 \kms$, which is roughly
the depth to which a galaxy with $M_B=-17$ can be seen given the magnitude
limit of the survey.  As expected, the density field is highly non-uniform on
this scale.  Two features of the galaxy distribution are particularly relevant to 
our measurement of the faint end of the LF.
First, the dominant feature at very low redshift is the void in
SSRS2 South (22$^h$-3$^h$, $cz \le 1,500 \kms$ in Figures 4$b$ and $d$).  Because of this underdensity,
the number of galaxies with $M_B \ge -15.5$ is quite small (48 galaxies with
$cz \ge 500 \kms$).
The second feature is the obvious overdensity between 3$^h$ and 4$^h$ in Figure 4$b$.
panel ($cz \approx 1,100 \kms$).  This region includes the Fornax and Eridanus
clusters, where virial motions are large and therefore individual galaxy distances are  
uncertain.
If a significant fraction of the intrinsically faint galaxies
are bound to these clusters, then we may expect some systematic error in the
faint end of the LF both from the incoherent velocity fields within the
clusters ({\it} i.e. the redshift fingers) and from the coherent streaming
motions induced by the large-scale density fluctuations.

In order to evaluate the effects peculiar motions, we explore two cases.
First, we compute the LF using a simple model for the local flowfield: 
we assume spherical infall to the Virgo cluster with $v_{inf}=250
\kms$ (case 1).  Although this case clearly ignores very local flows such as 
infall to Fornax, it serves as a good starting point.  In the second case, we again
assume spherical infall to Virgo but add 
the somewhat extreme assumption that all galaxies lying within 1.5 $\hmpc$ (projected) and $ 2,500 \kms$
of the centers of known clusters actually lie at the central redshift
of the clusters.  Although these cases are not exhaustive, they
should give us some idea of the degree to which peculiar
velocities affect our conclusions.  

Figure 5 shows the LFs computed in cases (1) and (2) along with the original
LFs shown in Figure 2.  For reference, the fitted Schechter functions are recorded in
the last eight rows of Table 1.
Although the general result of the Virgocentric flow correction
is to make galaxies brighter, the effects on the shape of the overall luminosity function are
quite small (panel $a$).  This result is not surprising: since the mass concentration driving
the flow sits in the northern hemisphere, the flow in the SSRS2 region
is essentially a bulk motion where all galaxies move in roughly the
same direction.  A small inflection appears at $M_B=-15.5$ after the correction, and the
number in the very faintest bin decreases by approximately a factor
of two (roughly the size of the original $1\sigma$ error bar).
Because the spirals are well dispersed through the
volume, it is also not surprising that the effects on the spiral LF are
small (panel $c$).  A similar inflection appears at $M_B=-15.5$, but again the
new LFs are consistent with the original estimates to within the $1\sigma$
errors.  Because of the presence of the clusters, one might expect the
case (2) corrections to have the greatest impact on the E/S0 LF.  However,
panel ($b$) shows that even the extreme assumption that all galaxies
near Fornax and Eridanus are in the cluster cores does not change the
LF significantly.  Finally, because most irregular/peculiar galaxies
are faint, one also expects their LF to be particularly sensitive
to local flows.  Once again, the changes in the LF are small.  The 
case of pure Virgocentric flow reduces the number of irregular/peculiar
galaxies brighter than $L_*$ somewhat, but overall, the luminosity
functions are all consistent.  The steep Schechter function is a reasonable 
approximation to the irregular/peculiar LF no matter what we assume for
the local flow field: the shallowest LF (case 1) has
a faint-end slope $\alpha=-1.74 \pm 0.25$.  The slope appears to be somewhat
flatter if we consider only galaxies fainter than $M_B=-18$, but even
in this magnitude range, the slope is still $\alpha \approx -1.5$.  We show
a Schechter function with this shallower slope as the dashed line in Figure 5.
Although this function is a very poor fit to the rest of the luminosity function,
it serves as a representative lower limit to the slope at the very faint end.  
We conclude that our luminosity functions depend only
weakly on the details of the local flow field.

The SSRS2 LFs closely resemble the LFs derived for similar morphological classes
in the CfA Survey (Marzke \etal 1994a) but disagree in some cases with those derived from the 
Stromlo-APM (Loveday \etal 1992).
In the SSRS2, the faint end of the Irr/Pec LF is much steeper than in any other class of galaxies.
Our best estimate of the slope is $\alpha=-1.81 \pm 0.24$, quite similar to
earlier measurements from the CfA: $\alpha=-1.88 \pm 0.2$.
As with the other morphological classes, however, the Irr/Pec
$M_*$ is considerably brighter in the SSRS2 than in the CfA
Survey, and $\phi_*$ is consequently lower.  In Figure 6,
a direct comparison of the luminosity functions shows
that the faint ends of the Irr/Pec LFs match
remarkably well; the difference between the two stems
from a decrement of bright galaxies in the CfA survey (or
possibly an excess in the SSRS2).  Unfortunately, irregular and peculiar
galaxies were combined with spirals in Loveday \etal (1992), and 
we cannot compare our results to the Stromlo-APM directly.

The LF of early-type galaxies in the SSRS2 is essentially flat
between $M_*$ and $M_{B}= -14$: $\alpha=-1.00\pm0.09$. At faint absolute
magnitudes, the  E/S0 LF
significantly exceeds the Stromlo-APM LF (Loveday \etal 1992) 
but is similar to the CfA LF (see Figure 6).  Again, however, the CfA LF shows a 
deficit of bright E/S0's compared to the SSRS2. 
The overall normalization of the CfA E/S0 LF is also somewhat higher than the SSRS2 even at the
faint end.
This excess of CFA E/S0s is consistent 
with the enhanced abundance of clusters in the CfA sample, most of which are concentrated
in the northern Galactic cap (Ramella \etal 1997, Marzke \etal 1995).

The deficit of bright galaxies in the CfA LFs appears consistently 
in each morphological bin and probably signals a systematic error in the 
Zwicky system.  The source of the discrepancy at the bright end between the CfA survey and nearly
every other survey is unclear; systematic errors in the Zwicky magnitude
scale remain poorly constrained.  
One might suppose that saturation in the Schraffierkassette films 
used in the Zwicky Catalog (on which the CfA Survey is based)
could lead to underestimation of the flux in bright galaxies.  
However, intrinsically bright galaxies appear over a wide range of 
{\it apparent} magnitudes in the CfA sample, and it is not immediately
clear that saturation could cause the observed depression at the bright end of
the LF.  The CfA LF could also be forced to agree with 
the SSRS2 by convolving an extra Gaussian error distribution with $\sigma \approx 0.5$mag
(in addition to the 0.35 magnitude error computed by Huchra (1976) and already
deconvolved during the computation of the LF).
In this case, however, it is not clear where such a large dispersion
would arise; random errors in the SSRS2 and CfA magnitudes are quite similar.
A definitive resolution of this problem
awaits the completion of wide-angle CCD surveys (e.g. Gunn 1995),
which will provide more accurate magnitudes for 
CfA (and maybe later SSRS2) galaxies and will provide a more detailed
understanding of the completeness of local redshift catalogs.

\section {Reconciling Galaxy Counts with the $z=0$ LF}

Three aspects of the local luminosity function are particularly relevant to 
the interpretation of intermediate-redshift morphology.
First, the overall normalization of the present-epoch LF affects the predicted galaxy counts
for all morphological types and is an important factor in the debate over very recent
galaxy evolution.
Second, the faint-end slope of the E/S0 LF plays a critical role in the debate over the
age and formation history of early-type galaxies.  Because the deep number counts
of early-type galaxies differ least from no-evolution predictions, the shape of the local
LF has the largest effect on the inferred evolution.  Finally, the abundance of local
irregulars and the shape of the irregular-galaxy LF are important to our understanding
of the remarkable irregularity observed in the faint blue galaxy population.
In the following sections, we focus on each of these aspects individually.

\subsection {Normalization}

Maddox \etal (1990) were the first to draw attention to the remarkably bright apparent
magnitude at which the observed galaxy counts diverge from no-evolution predictions.
They interpreted their results as evidence of very recent evolution in the galaxy population.
Since then, the steep galaxy counts at $B \le 20$ have been ascribed to a number of alternative sources
including large-scale density fluctuations, scale errors in the APM magnitudes and 
systematic errors in the APM galaxy detection.  Given the uncertainty in the plate
surveys, is it reasonable to predict faint galaxy counts using luminosity
functions derived from bright galaxy redshift surveys?  In this section,
we attempt to address this question quantitatively.

Although we can measure the {\it shape} of the luminosity function accurately even in the
presence of large density fluctuations, our {\it normalization} is fixed by the galaxy
counts to our limiting magnitude and is therefore sensitive to local density
anomalies.  Large-scale structure affects the counts when galaxies
cannot be seen to distances much larger than the size of the structure.  At $B=15.5$,
the observable redshift range is only a few times the typical void size.  
At even brighter magnitudes, we expect the observed counts to exceed the predictions 
of a homogeneous model simply because our viewpoint is not a random one: other galaxies
are correlated with our own.  

Figure 7 shows the SSRS2 galaxy counts (solid triangles) along with the predictions based on the 
overall luminosity function from Table 1, k-corrections for an Sb galaxy
(an appropriate mean for a $B$-selected sample at this depth),
our chosen cosmological parameters: $H_0=100, q_0=0.2$, and an assumed dispersion
in the photometry of 0.3 magnitudes (solid line).  Because the normalization of the LF is largely determined
by galaxies fainter than $B=14.5$, it is not surprising that the predicted
counts agree well with the observations at these magnitudes.  Even so, fluctuations caused
by large-scale structure are apparent even at $B=15.0$; in this case, the excess in the counts
is caused by the overdensity of galaxies at approximately 6,000 $\kms$ previously labeled
the Southern Wall (da Costa \etal 1998; also see Figure 4).  
As expected, local clustering also flattens the slope of the observed counts at
brighter magnitudes.  At the magnitude limit of the survey, the agreement between
the observed and predicted counts is excellent.

The open circles in Figure 7 represent the APM counts from Maddox \etal (1990).
As they pointed out, the counts depart from the no-evolution predictions at
approximately $B=17$.  It is worth pointing out that 
in the small region of overlap ($B \approx 15.5$), the SSRS2 counts reproduce the APM
counts very well.  The construction of the SSRS2 photometric catalog does not 
force this agreement; initial candidates for the SSRS2 included
all sources flagged as non-stellar in the STScI scans of the SERC $J$ plates.  Although both
surveys share the same original plate material, the STScI survey is based on PDS scans
with a higher dynamic range, and algorithms for detection and photometry were 
constructed independently.

Given the recent reports of possible scale errors in the APM (Metcalfe \etal 1995, Bertin \&
Dennefeld 1997$a$), the agreement between the APM and the SSRS2 (which is independently calibrated 
using extensive CCD photometry) may seem surprising.  However, the reported problems in the APM 
appear primarily at fainter magnitudes than are probed by
the SSRS2: $16 \le B_J \le 18$.  Metcalfe \etal (1995) claim that the low APM counts
at magnitudes brighter than this (and therefore in the range of the SSRS2) 
cannot be explained by the same errors.  On the other hand, 
Bertin \& Dennefeld (1997$a$) attribute their lower counts at $B_J \le 16$ to incompleteness in their
photographic catalog, and they caution that similar incompleteness affects other photographic
surveys as well.  They identified star/galaxy separation as the primary culprit behind
their incompleteness at the bright end.  It is important to point out, however, 
that the SSRS2 grew out of the nonstellar
sources of the STScI Guide Star Catalog, where the definition of ``nonstellar'' is necessarily 
conservative. Because the astrometric requirements of the HST were the primary consideration in
the construction of the GSC, 
Lasker \etal (1990) tuned the stellar classifier to assure a clean {\it stellar}
catalog. Even so, roughly 2\% of the SSRS2 galaxies were originally classified 
as stars in the GSC but were labeled galaxies in the APM, ESOLV or MCG (da Costa \etal 1998).  
This fraction is consistent with the success rate established by Lasker \etal (1990) using
visual checks and multiply observed sources.
Because stars at this apparent magnitude are very numerous compared to galaxies, it is of course
possible that some galaxies were left for stars in all four catalogs and therefore did not make 
it into the final version of the SSRS2.  

The number of very compact galaxies missed in the bright galaxy counts
can only be resolved by blind redshift surveys of all detected sources in a field.  At these 
magnitudes, the large ratio of stars to galaxies is a major obstacle. 
Although we lack strong observational constraints, 
we can roughly gauge the magnitude of the problem using theoretical
predictions of the distribution of galaxy sizes.  Such predictions
relate the surface brightness profiles and rotation curves of galaxy disks to the 
distributions of mass and angular momentum in the parent dark-matter halos (Dalcanton, Spergel 
\& Summers 1997, Mo, Mao \& White 1997).  Because these models are reasonably successful
at matching the properties of galaxies we see, they may provide some useful clues 
about those we don't.  Here, we are particularly interested in the distribution
of scale lenghts at a fixed halo mass (in this case the mass corresponding to an $L_*$
galaxy, $\sim 10^{12} M_\odot$), which depends only on the distribution of angular momenta.
Using reasonable parameters for the distribution of halo angular momenta and the baryon
fraction as outlined in Dalcanton \etal (1997),
we find that the the fraction of $L_*$ galaxies with scale lengths less than
a kiloparsec (which corresponds roughly to the size of the point-spread function at the maximum depth of the SSRS2)
is smaller than 10\% if $\Omega_0=1$.  At lower $\Omega_0$, the fraction is even smaller.  Of course,
these models are only a first step and require further testing, but to first
order, we expect a very small error in the density of $L_*$ galaxies
from misclassification of compact galaxies.

At the other end of the surface-brightness spectrum, we also expect some bias against very diffuse
galaxies in the SSRS2.  
As originally suggested by Disney (1976), local surveys of low-surface-brightness (LSB) galaxies have shown that 
the range of surface brightness covered in traditional, ``magnitude-limited''
redshift surveys is limited (Sprayberry \etal 1997).
However, Sprayberry \etal also showed that the vast majority of galaxies overlooked in
surveys like the SSRS2 are intrinsically faint (see their Figure 4).  The contribution
of LSBs to the galaxy density at $L_*$ is less than 10\%. Because the galaxy counts are
largely determined by galaxies near $L_*$, (unless the luminosity function of LSBs is
much steeper than observed), the bias against LSBs is unlikely to affect the
counts significantly.  At $L_*$, we therefore expect the combined errors from
galaxy detection and star/galaxy separation to be smaller than 20\%.

If the steep slope cannot be blamed on errors in detection or photometry, then
the next most plausible alternatives are recent evolution in the galaxy population or
large-scale density fluctuations.  Both of these options have been constrained
somewhat by recent redshift surveys.  Ellis \etal (1996) have shown that the space density
of galaxies derived from the Autofib redshift survey is a factor of two larger than
we derive from the SSRS2 in the redshift range $0 \le z \le 0.1$.  However, it is
important to point out that 
most galaxies used in their computation lie near the upper limit of the redshift
bin, $z=0.1$.  Furthermore, Ellis \etal eliminated the DARS (Peterson \etal 1986)
from the analysis of the Autofib survey when they concluded
that the space density of galaxies in this sample was anomalously low.
The magnitude limit of the DARS is $B=17$, comparable to other surveys finding
low normalizations.
The exclusion of the DARS was well justified: because Ellis \etal (1996) 
computed the LF using a technique that is biased by density fluctuations,
they recognized that the {\it shape} of the LF computed from the combined (bright+faint) samples would
be biased if the overall densities in the individual samples were significantly different.  
However, the elimination of low-redshift, low-normalization
regions clearly biases the determination of the overall normalization at $z \le 0.1$.

Using the field complement of the Norris Survey, Small, Sargent and Hamilton (1997) 
also measure a large space density 
(similar to Autofib) at $z \le 0.2$.  In this case, the mean redshift of the low-redshift
sample is $z=0.15$, and once again the survey is essentially disjoint from the SSRS2, the
Stromlo-APM, and other low-normalization surveys.  The Century Survey  (Geller \etal 1997),
which samples the entire region $z \le 0.15$, yields evidence of a 50\% increase in the 
normalization at $z \ge 0.06$. Geller \etal note that errors in the
selection function may mimic such an increase, and they interpret their data cautiously.
Taken together, however, these surveys
provide constraints on the timescale (or alternatively the spatial scale) over which
the normalization would have to change.
The luminosity function appears relatively stable between $z=0.2$ and
$z=0.1$ and then drops rapidly at lower redshift.  In order to match the density field
measured from the low-redshift surveys, the change must be remarkably
abrupt; these surveys indicate little change in the density at $z \le 0.1$ (\eg Loveday
\etal 1992, Ratcliffe \etal 1998).  Such a
discontinuity seems very unlikely to be an evolutionary effect, especially given that
the galaxy counts in the $K$-band show a similar trend (Huang \etal 1996, but see Bertin \& Dennefeld 1997$b$
for a dissenting view).  Unless there are large, systematic 
errors swaying all of the low-redshift surveys in the same direction, then the most
likely explanation is that a very large portion of the southern sky (at least $150 \hmpc$ across)
is underdense by approximately a factor of two.

If a local hole is the source of the anomalously steep galaxy counts, then it must be remarkably uniform
over the region covered by the SSRS2.  We see no evidence of substantial density gradients
from one part of the survey to another.  Figure 8 shows a comparison between the SSRS2 subsamples
in the northern and southern Galactic hemispheres.  In this figure, we show the redshift distribution
for each subsample along with the predictions based on a single luminosity 
function computed from the entire SSRS2.  The solid curve in panels $a$, $c$ and $e$ represents the
expected redshift distribution for a uniform distribution of galaxies in space.  Each curve is
scaled only by the solid angle of the subsample.  Large-scale structure is clearly evident
in each panel; however, the overall density of galaxies is remarkably constant from one
section of the survey to the other.  To check this, we have also fitted luminosity functions
to each subsample independently.  The luminosity functions for SSRS2 North and South are
consistent at the 1$\sigma$ level in both shape and normalization, and they are also 
indistinguishable from the overall SSRS2 luminosity function.  This agreement is reflected
in the shape and the overall normalization of the redshift distributions in Figure 8.
The most extensive feature in either subsample is the underdensity in the north
at $z \approx 0.035$. Although this underdense region is quite large, it is still
counteracted by the overdensities at lower redshift in such a way that the overall
density of SSRS2 North is consistent with the SSRS2 as a whole.

Figure 8 also shows very little trend in the observed mean density with redshift.
Panels $b$, $d$ and $f$ show the ratio of the observed redshift distribution
to the expected distribution, and aside from the well-known fluctuations caused by
typical voids and walls, the ratio is remarkably stable.  The error bars in this panel
have been left out for clarity, but it is important to note that beyond $z=0.05$, the ratio 
is highly uncertain (as can be seen from the small number of galaxies in the corresponding
panel $a$).  Tiny uncertainties in the shape of the luminosity function also contribute
to large uncertainties at $z \ge 0.05$.  Although there appears to be some hint of a
density increase in this redshift range, it is not statistically significant in either
subsample or in the SSRS2 as a whole.

According to our current picture of large-scale structure, a local underdensity of the required
magnitude and covering such a large volume would be surprising; the {\it rms} fluctuations
computed from the observed power spectrum are only $\sim$10\% on this scale (\eg Baugh 1996).  
If density fluctuations are Gaussian, then the probability of observing a factor of two underdensity
in our own backyard seems disturbingly small. 
However, recent progress in 
our understanding of large-scale structure
has been punctuated by a few notable surprises 
(\eg de Lapparent, Geller \& Huchra 1986, Davis \etal 1982), and it seems premature 
to dismiss the possibility of very large-scale structure on this basis alone.  
Recent observations suggest that such large density fluctuations may not be so rare:
the Corona Borealis region, for example,  exceeds the mean galaxy density
by $\sim$70\% on scales of $100 \hmpc$ (Geller \etal 1997).
Even a glance at the LCRS redshift distribution suggests that at least
some modulation of the density on these very large scales is common.
More quantitatively, Landy \etal (1996) have found excess power in the two-dimensional
power spectrum at $\sim 100 \hmpc$.  It remains to be seen whether the frequency
of such large-scale features is large enough to remove the novelty of the putative local hole.  
Surveys covering larger
volumes will be necessary to resolve this question definitively.

In an attempt to be as faithful to the entire range of observations as possible, 
we create a fiducial no-evolution model for the galaxy counts
which accounts for a low-redshift change in the normalization.
We form this prediction by assuming that the shape of the LF does not evolve but
that the density increases smoothly over the redshift range $0.08 \le z
\le 0.12$.  Some justification of this assumption can be found in the
LCRS, which covers the entire region $0 \le z \le 0.2$. 
In Figure 8 of Lin \etal (1996), there is some evidence
of a discontinuity in the density field at approximately $z=0.07$.  The magnitude of the
observed jump falls short of the required factor of 2.3, but the volume in this
region of the LCRS is quite small.  We have arbitrarily chosen a smooth function
to bridge the gap between the low-normalization and high-normalization regions:
$\phi_*(z)= \phi_*(z<0.05)f(z)$ where $f(z) =  1 +  \lbrace \exp{\lbrack-(z-z_c)/\Delta z\rbrack} + 1 \rbrace ^{-1}$ 
with $z_c=0.1$ and $\Delta z=0.01$.  This fiducial model of the total galaxy counts is 
shown as the dashed line in Figure 7 and is roughly consistent with the galaxy 
counts to $B\sim 20$.

\subsection {The Faint End of the E/S0 LF}

The spectacular progress in our ability to discern faint-galaxy morphology has
inspired a re-evaluation of the faint end of the local E/S0 luminosity function.
As we noted in \S4, the best available estimates of the field E/S0 LF are strongly contradictory.
Loveday \etal (1992) found a steeply decreasing faint end in the Stromlo-APM
survey ($\alpha=+0.2 \pm 0.35$), while Marzke \etal (1994$a$) found a flat faint end
($\alpha=-0.97 \pm 0.2$) in the CfA survey.  Both Marzke \etal (1994$a$) and Zucca \etal (1994)
showed that incompleteness in the Stromlo-APM survey could be
responsible for the steeply declining faint end.  Because only 1310 of the
1658 Stromlo-APM galaxies were morphologically classified, Marzke \etal (1994$a$) suggested that
the correlation between intrinsic luminosity and classifiability (essentially apparent size in this case) 
could have caused a bias against intrinsically faint, early-type galaxies.
Monte Carlo simulations showed that the bias was sufficient to cause the observed
discrepancy in the LF and also to account for the anomalously low $V/V_{max}$ 
obtained for this sample.  Because of the uncertainty in
both the Stromlo-APM LF and the CfA LF, the true abundance of faint, local E/S0's
has remained a nagging question.

With the arrival of the HST Medium Deep Survey (Griffiths \etal 1994,ApJ,437,67) and the Hubble Deep
Field (Williams \etal 1996), the possibility of a flat E/S0 LF 
gained new popularity.  Glazebrook \etal (1995) and Driver \etal (1995$b$)
showed that a flat E/S0 LF combined with a high overall normalization 
produced no-evolution predictions which very nearly matched the E/S0 counts from
the HST Medium Deep Survey.  Driver \etal (1995$a$) drew similar conclusions
from a very deep HST field near a high-redshift radio galaxy, and 
Abraham \etal (1996) used the HDF counts to follow the trend to $I=25$.
As we discussed in \S5.1, the case for a high normalization at modest redshifts
has been strengthened by the results of the Autofib, Norris and Century redshift surveys. 
However, the choice of $\alpha \approx 1$, which is critical to the conclusion that early-types
follow the no-evolution predictions, has remained somewhat arbitrary.

Because it combines the advantages of the Stromlo-APM and the CfA surveys,
the SSRS2 is uniquely capable of improving this measurement.
Like the Stromlo survey, the SSRS2 is based on a well-calibrated and reproducible
photometric catalog.  On the other hand, the SSRS2 targets nearby (and therefore
apparently large) galaxies and thus allows more detailed galaxy classification.
Like the CfA survey, morphological classification in the SSRS2
is more than 99\% complete.  The flat faint-end slope measured for early-types in the SSRS2
lends strong support to the conclusions of Driver \etal (1995$a,b$), Glazebrook \etal (1995)
and Abraham \etal (1996) regarding the counts of faint Es and S0s.  In order to 
provide a useful benchmark, we have computed no-evolution models based on our measured
LFs and recorded them in Tables 3 and 4.  These models include the redshift-dependent
$\phi_*$ discussed in \S5.1 and type-dependent K-corrections based on the Rocca-Volmerange
\& Guiderdoni (1988) models.  We compute $N(m)$ in $B$ and then in $I_{814}$ using a mean
color for each type from Windhorst \etal (1994): $\langle B-I \rangle$ = 2.3, 1.9 and 1.4 for
E/S0s, spirals and irregulars, respectively.  We include models for two values of 
the deceleration parameter: for each type, the first column represents $q_0=0.5$ while
the second column represents $q_0=0.05$.  The values listed in the tables are
raw galaxy counts {\it before} convolution with a magnitude-error distribution.
 
Figure 9 compares the HDF counts with the $I$-band no-evolution models.  These models,
which are intended to represent the most objective possible interpretation of
both the low-$z$ and intermediate-$z$ redshift surveys, reproduce the E/S0
counts very well.  Indeed, given that stars in these galaxies obviously
evolve with time, the agreement is almost {\it too} good; one expects
passive luminosity evolution to produce at least some enhancement in the
counts at faint magnitudes unless it is counteracted by a decrease
in number density.  Ongoing studies of the colors and redshift distributions
of these faint early-types should help to disentangle these competing effects.

Although the case for a flat E/S0 LF now seems quite convincing, the connection
to other LFs based on spectral type and color remains somewhat puzzling.
For example, Lin \etal (1996) have
shown that galaxies with weak OII emission have a steeply declining LF ($\alpha=-0.3
\pm 0.1$) at magnitudes
fainter than $M_*$.  Given the correlation between star-formation and morphology,
this seems surprising; one expects the weak-OII LF to reproduce the E/S0 LF at least
to first order.  A more detailed analysis of the LCRS confirms the dependence of the LF
on spectral type and shows a continuous variation from a steeply declining LF
for the earliest spectral types to steeply increasing LFs for 
star-forming galaxies (Bromley \etal 1997).  A similar picture emerges from the
dependence of the SSRS2 LF on color:
the LF of red galaxies in the SSRS2 declines somewhat at the faint
end ($\alpha=-0.73 \pm 0.24$) while the blue galaxy LF is relatively steep
($\alpha=-1.46 \pm 0.18$, Marzke \& da Costa 1997).  The fact that the color dependence mimics
the dependence on emission strength seems reasonable: both are strongly tied
to recent star formation, while the processes governing galaxy morphology
are less clear.  The large scatter in the relations between color and morphology
and between line strength and morphology may hide strong luminosity dependencies, which
in themselves would provide interesting constraints on galaxy formation.
Samples large enough (and homogeneous enough) to determine the full multivariate
distribution in luminosity, color, morphology and spectral type
will clearly provide a major step forward in our understanding of galaxy
formation.

As a final note, we emphasize that we do not distinguish in this paper between the 
various classes of spheroidal galaxies.
Dwarf spheroidals, giant ellipticals and lenticular galaxies 
all fall into our E/S0 bin.  Using these coarse morphologies
alone, we obviously cannot comment on the LFs of more specific classes
of galaxies.  For example, the very detailed morphological decomposition of the Virgo cluster
luminosity function (Sandage, Binggeli \& Tammann 1985)
reveals a series of roughly Gaussian type-specific luminosity functions with
varying central magnitudes; only the dEs are unbounded at the faint end
(to the limit of the survey, at least).
Upon closer examination, the field E/S0 LF may reveal a similar construction.
One of us is currently obtaining CCD imaging of faint SSRS2 galaxies to explore this possibility further.

\subsection {The Shape of the Irregular/Peculiar LF}

The remarkable increase in irregular morphology at faint apparent magnitudes
reported by Driver \etal (1995$a,b$), Glazebrook \etal (1995) and Abraham \etal (1996)
raises two related questions: first, what fraction of
these faint irregular galaxies are simply nearby dwarfs?
Second, if the majority of these irregulars lie at high redshift, what are their
descendents at the present epoch?

Koo, Gronwall \& Bruzual (1993) showed that uncertainties in the
local luminosity function (particularly at the faint end) play a pivotal
role in our interpretation of the faint galaxy counts.
If our tally of nearby dwarfs is incomplete, then no-evolution
predictions will fall short of the observed galaxy counts even in the absence 
of real evolution.  Marzke, Huchra \& Geller (1994$b$) showed that 
such an underestimate had indeed occurred: at the faint end of the
luminosity function, there were more galaxies than would be 
predicted by the extrapolation of the Schechter function fitted
to the bright end.  However, the magnitude of the excess was 
uncertain, and because of the small volume of the universe probed
at these faint absolute magnitudes, it remained unclear whether this
excess was peculiar to the local region or whether it was a global
feature of the present epoch.  

Later, Marzke \etal (1994$a$) determined that the galaxies responsible for the 
faint-end excess were very late-type spirals and irregulars.
The luminosity function of this class was remarkably steep: $\alpha=-1.88 \pm 0.2$.
As we discussed earlier, however, the Zwicky magnitude scale
was uncertain, and the overall contribution of these nearby late types
to the deep galaxy counts was unclear.  
The improved photometric scale and the well-defined
detection algorithm used in the construction of the SSRS2 allow us to 
measure the irregular-galaxy LF with much greater confidence.  The slope of
the Irr/Pec LF we measure here is essentially identical to the slope of
the CfA Sm-Im LF: $\alpha=-1.81\pm0.24$.  Although the random error
associated with the new measurement is comparable to the older CfA
measurements, the systematic uncertainties, which were difficult
to quantify in the CfA, are certainly reduced in the SSRS2.
Nevertheless, Figure 6 shows that the faint end of the Irr/Pec LF reproduces
the CfA Sm-Im LF very well.  As noted in \S4, the differences between the two LFs occur
at the bright end.  

As with the CfA survey, however, the 
volume surveyed at the faintest absolute magnitudes is still quite small, and if the shape
of the Irr/Pec LF depends on the particular details of the very nearby
universe, then it may be unwise to extrapolate this result to the faint
galaxy counts.  However, several recent observations suggest that the
steep Irr/Pec LF is not a local oddity.  Bromley \etal (1997) have shown
that the class of LCRS galaxies exhibiting the most active star formation
(according to their spectral energy distributions) have a very steep luminosity 
function: $\alpha=-1.93\pm0.13$.  Although the LCRS sample does not have detailed
morphological classifications, the fact that essentially all irregular galaxies
show strong emission lines suggests that these LFs
represent the same class of galaxies.  Although the LCRS LFs do not extend
as faint as the CfA or SSRS2, their redshift range is much greater, and
the possibility that the steep LF is a local fluke seems less likely.

Another related observation is the recent UV-selected survey of Treyer \etal (1996).
This survey of $\sim 40$ galaxies covers the redshift range $0 \le z \le 0.3$ and
is selected by 2000$\AA$ flux.  Because of the huge variation in $m_{2000}-B$ across
the Hubble sequence, this sample is dominated by late-type, star-forming galaxies.
Although the sample is small, Treyer \etal were able to show that the faint-end
slope of the LF is steep: $\alpha=-1.77 \pm 0.15$, clearly consistent with the
Irr/Pec LF measured here, the LF of star-forming galaxies measured by Bromley
\etal (1997), the luminosity function of the bluest quartile in the SSRS2 ($\alpha \approx -1.7$, Marzke
\& da Costa 1997) and the Sm-Im LF of the CfA survey  (Marzke \etal 1994).
It is also worth noting that the observed HI mass function of gas-rich galaxies is not
inconsistent with an upturn at the low-mass end; the distribution is still poorly determined
in this regime (Schneider 1997, Zwaan \etal 1997).
Given the growing consensus, it now seems reasonable to conclude that the
steep luminosity function is a universal feature of star-forming galaxies.

The last panel in Figure 9 shows the HDF counts of irregulars and peculiars along with
the no-evolution predictions described in \S5.2.  Although our models differ
in detail with the earlier models from Driver \etal (1995$a,b$), Glazebrook \etal (1995) and Abraham
\etal (1996), the excess of faint irregulars is so pronounced that the overall conclusion
remains unchanged: the no-evolution models cannot account for the irregular population at faint magnitudes.  
It appears very unlikely that remaining uncertainties in the local luminosity function will be able to make up 
the difference.  For example, the two sets of curves represent different
assumptions about the behavior of the LF at fainter luminosities than we can measure from 
the SSRS2: the lower curve represents a cutoff at $M_B=-14$, the limit of our
survey, while the upper curve extrapolates the steep LF all the way to $M_B=-10$.  
The extrapolation makes very little difference.

More detailed information about the faint irregular population is slowly
becoming available.  Perhaps the most robust conclusion that can be drawn from a comparison
of our local survey to the results of the deep redshift surveys (Brinchman \etal 1997, Lilly \etal 1997)
is that the super-$L_*$ irregulars observed by Brinchman \etal (1997) and Glazebrook \etal (1997) 
at redshifts greater than a half have very few local counterparts.
For example, the density of galaxies at $M \approx -21$ is more than an order of magnitude
higher at $z \ge 0.75$ than it is locally.  Brinchman \etal drew a similar conclusion from their own data, which
yielded no super-$L_*$ irregulars at $z \le 0.5$ even though they 
could have been detected had they been present in the numbers observed at higher redshift.
With larger, more homogeneous surveys planned for the near future,
the relationship between the bright irregulars at high redshift
and their faint, low-redshift counterparts should soon become more clear.

\section{Conclusions}

We have used a new sample of 5,404 galaxies with rough but complete morphological classifications
to determine the galaxy luminosity function for different morphologies over the range $-22 \le M_{B}
\le -14$.  We conclude that the faint-end slope for both E/S0 and spiral galaxies is essentially flat over
this range.  The LF of irregular and peculiar galaxies 
is very steep ($M_*-19.78^{+0.40}_{-0.50}+5\log{h}$, $\alpha=1.81^{+0.24}_{-0.24}$,
$\phi_*=0.2 \pm 0.08 \times  10^{-3} h^3\,
\rm{Mpc}^{-3}$). The faint-end slope of the Irr/Pec LF is consistent with earlier
measurements from the CfA Redshift Survey, however, an excess of bright irregulars
relative to the CfA LF leads to a brighter value of $M_*$ for this class in the SSRS2.
This pattern appears in each morphological class and may be evidence that bright galaxies
are generally underrepresented in the CfA Survey.

The flat faint-end slope of the E/S0 LF supports earlier claims that the Stromlo-APM
LF under-represents faint early-type galaxies.  As a result, the no-evolution predictions
of faint E/S0 counts based on the SSRS2 exceed the predictions based on the Stromlo-APM
and, assuming that the high normalization obtained for the intermediate-redshift LFs is
representative, the SSRS2 predictions are consistent with the observed counts of Es and S0s to $I=25$.
As with other surveys to similar depths, however, the normalization obtained directly from
the SSRS2 is low, and the 
explanation of this low-redshift anomaly remains elusive.
Until the biases in the present-epoch luminosity function are better understood, 
the degree of evolution inferred from deep counts will remain uncertain.

\acknowledgments

We thank Todd Small, Huan Lin, Julianne Dalcanton, Rebecca Bernstein, Elena Zucca and
Richard Ellis for helpful discussions.  We also thank Seth Cohen for his help with
Figure 7 and an anonymous referee for stimulating a more lengthy version of \S5.

\clearpage

\begin{center}
{\bf TABLE 1}

Schechter Function Parameters \\ $-22 \le M_B \le -13.5$ \\
\vskip 0.1in
\begin{tabular}{ccccccc}
\hline
\hline
  &Sample  & $N_{gal}$ & $M_*$  &$\alpha$ 
  &$\phi_* ({10^{-3} \,\rm Mpc^{-3})}$ &$P(\ln{\cal L}_1/{\cal L}_2) $ \\
\hline  
&&&&&&\\
&All & 5036 & $-19.43^{+0.06}_{-0.06}$ & $-1.12^{+0.05}_{-0.05}$ &12.8$\pm$2.0 & 0.03 \cr
&&&&&&\\
&E/S0 & 1587 & $-19.37^{+0.10}_{-0.11}$ & $-1.00^{+0.09}_{-0.09}$ &4.4$\pm$0.8 & 0.04 \cr
&&&&&&\\
&Spirals & 3227 & $-19.43^{+0.08}_{-0.08}$ & $-1.11^{+0.07}_{-0.06}$ &8.0$\pm$1.4 & 0.03 \cr
&&&&&&\\
&Irr/Pec & 204 & $-19.78^{+0.40}_{-0.50}$ & $-1.81^{+0.24}_{-0.24}$ &0.2$\pm$0.08 & 0.16 \cr
&&&&&&\\
\hline
&&&&&&\\
 &All & 5054 & $-19.40^{+0.06}_{-0.06}$ & $-1.08^{+0.06}_{-0.05}$ &13.7$\pm$1.9 & 0.02 \\
&&&&&&\\
Case 1 &E/S0 & 1592 & $-19.38^{+0.10}_{-0.11}$ & $-0.99^{+0.10}_{-0.09}$ &4.5$\pm$0.9 & 0.03 \\
Infall&&&&&&\\
 &Spirals & 3240 & $-19.38^{+0.08}_{-0.08}$ & $-1.06^{+0.07}_{-0.07}$ &8.8$\pm$1.2 & 0.02 \\
&&&&&&\\
 &Irr/Pec & 204 & $-19.69^{+0.39}_{-0.48}$ & $-1.74^{+0.25}_{-0.25}$ &0.2$\pm$0.1 & 0.09 \\
&&&&&&\\
\hline
&&&&&&\\
 &All & 5060 & $-19.41^{+0.06}_{-0.06}$ & $-1.09^{+0.06}_{-0.05}$ &13.5$\pm$1.9 & 0.02 \\
&&&&&&\\
Case 2 &E/S0 & 1595 & $-19.37^{+0.10}_{-0.11}$ & $-1.00^{+0.10}_{-0.09}$ &4.5$\pm$0.9 & 0.02 \\
Infall&&&&&&\\
 &Spirals & 3243 & $-19.38^{+0.07}_{-0.08}$ & $-1.06^{+0.07}_{-0.07}$ &8.7$\pm$1.2 & 0.03 \\
&&&&&&\\
 &Irr/Pec & 204 & $-19.74^{+0.40}_{-0.49}$ & $-1.79^{+0.24}_{-0.25}$ &0.2$\pm$0.1 & 0.06 \\
&&&&&&\\
\hline
\end{tabular}
\end{center}

\clearpage
\hoffset=-0.5in
\begin{center}
{\bf TABLE 2}

SWML Luminosity Functions \\
(log number $\rm Mpc^{-3}\,mag^{-1}$) \\
\vskip 0.1in
\begin{tabular}{ccccccccc}
\hline
\hline
&\multicolumn{2}{c}{All}&\multicolumn{2}{c}{E/S0}&\multicolumn{2}{c}{Spirals}&\multicolumn{2}{c}{Irr/Pec}\\ \cline{2-3} \cline{4-5} \cline{6-7} \cline{8-9}
Magnitude& $N$& $\phi(M)$ & $N$ & $\phi(M)$ & $N$ &$\phi(M)$ & $N$& $\phi(M)$ \\
\hline
&&&&&&&&\\
-21.10&239&$-3.60^{+0.03}_{-0.03}$&75&$-4.12^{+0.05}_{-0.05}$&157&$-3.79^{+0.03}_{-0.04}$&7&$-5.23^{+0.15}_{-0.23}$\\
&&&&&&&&\\
-20.32&1208&$-2.78^{+0.01}_{-0.01}$&387&$-3.27^{+0.02}_{-0.02}$&785&$-2.98^{+0.02}_{-0.02}$&32&$-4.47^{+0.09}_{-0.12}$\\
&&&&&&&&\\
-19.53&1525&$-2.31^{+0.01}_{-0.01}$&493&$-2.79^{+0.02}_{-0.02}$&978&$-2.53^{+0.01}_{-0.01}$&51&$-3.91^{+0.08}_{-0.09}$\\
&&&&&&&&\\
-18.74&1007&$-2.07^{+0.02}_{-0.02}$&320&$-2.58^{+0.03}_{-0.03}$&655&$-2.27^{+0.02}_{-0.02}$&28&$-3.65^{+0.09}_{-0.11}$\\
&&&&&&&&\\
-17.94&517&$-1.91^{+0.02}_{-0.02}$&161&$-2.42^{+0.04}_{-0.04}$&316&$-2.15^{+0.03}_{-0.03}$&36&$-2.98^{+0.08}_{-0.10}$\\
&&&&&&&&\\
-17.16&237&$-1.86^{+0.03}_{-0.04}$&67&$-2.43^{+0.06}_{-0.07}$&153&$-2.07^{+0.04}_{-0.05}$&17&$-2.84^{+0.12}_{-0.16}$\\
&&&&&&&&\\
-16.36&174&$-1.71^{+0.04}_{-0.04}$&47&$-2.33^{+0.07}_{-0.09}$&108&$-1.92^{+0.05}_{-0.06}$&18&$-2.52^{+0.13}_{-0.18}$\\
&&&&&&&&\\
-15.58&83&$-1.70^{+0.05}_{-0.06}$&26&$-2.31^{+0.09}_{-0.11}$&46&$-1.94^{+0.07}_{-0.08}$&9&$-2.48^{+0.16}_{-0.25}$\\
&&&&&&&&\\
-14.78&23&$-1.55^{+0.09}_{-0.12}$&6&$-2.35^{+0.16}_{-0.26}$&14&$-1.68^{+0.12}_{-0.16}$&3&$-2.17^{+0.24}_{-0.58}$\\
&&&&&&&&\\
-14.00&10&$-1.18^{+0.14}_{-0.21}$&1&$-2.30^{+0.31}_{-\infty}$&6&$-1.36^{+0.18}_{-0.30}$&3&$-1.40^{+0.30}_{-1.63}$\\
&&&&&&&&\\
\hline
\end{tabular}
\end{center}

\clearpage

\begin{center}
{\bf TABLE 3}

$B$-band No-evolution Models with a Local Hole \\
(log counts deg$^{-2}$ mag$^{-1}$)
\vskip 0.1in
\small
\begin{tabular}{ccccccccccccc}
\hline
\hline
   $B$ &&\multicolumn{2}{c}{All}&&\multicolumn{2}{c}{E/S0}&&\multicolumn{2}{c}{Spiral}&&\multicolumn{2}{c}{Irr/Pec} \\
\hline
10.5 && -2.86 & -2.86 && -3.36 & -3.36 && -3.06 & -3.06 && -4.28 & -4.28 \\
11.0 && -2.56 & -2.56 && -3.06 & -3.06 && -2.77 & -2.77 && -3.98 & -3.98 \\
11.5 && -2.27 & -2.27 && -2.77 & -2.77 && -2.47 & -2.47 && -3.68 & -3.68 \\
12.0 && -1.97 & -1.97 && -2.48 & -2.48 && -2.18 & -2.18 && -3.38 & -3.38 \\
12.5 && -1.68 & -1.68 && -2.19 & -2.19 && -1.89 & -1.89 && -3.09 & -3.09 \\
13.0 && -1.39 & -1.39 && -1.90 & -1.90 && -1.60 & -1.60 && -2.79 & -2.79 \\
13.5 && -1.10 & -1.10 && -1.61 & -1.61 && -1.31 & -1.31 && -2.50 & -2.50 \\
14.0 && -0.82 & -0.82 && -1.33 & -1.33 && -1.02 & -1.02 && -2.20 & -2.20 \\
14.5 && -0.54 & -0.54 && -1.05 & -1.05 && -0.74 & -0.74 && -1.91 & -1.91 \\
15.0 && -0.26 & -0.26 && -0.77 & -0.77 && -0.47 & -0.47 && -1.63 & -1.63 \\
15.5 &&  0.01 &  0.01 && -0.50 & -0.50 && -0.20 & -0.19 && -1.34 & -1.34 \\
16.0 &&  0.28 &  0.28 && -0.24 & -0.24 &&  0.07 &  0.07 && -1.06 & -1.06 \\
16.5 &&  0.54 &  0.54 &&  0.02 &  0.02 &&  0.33 &  0.33 && -0.78 & -0.78 \\
17.0 &&  0.80 &  0.80 &&  0.28 &  0.28 &&  0.59 &  0.60 && -0.50 & -0.50 \\
17.5 &&  1.07 &  1.07 &&  0.54 &  0.55 &&  0.87 &  0.87 && -0.21 & -0.21 \\
18.0 &&  1.35 &  1.35 &&  0.82 &  0.82 &&  1.15 &  1.15 &&  0.08 &  0.08 \\
18.5 &&  1.62 &  1.63 &&  1.09 &  1.10 &&  1.42 &  1.42 &&  0.36 &  0.37 \\
19.0 &&  1.88 &  1.89 &&  1.35 &  1.36 &&  1.68 &  1.68 &&  0.64 &  0.64 \\
19.5 &&  2.12 &  2.13 &&  1.58 &  1.60 &&  1.91 &  1.93 &&  0.91 &  0.91 \\
20.0 &&  2.34 &  2.36 &&  1.80 &  1.82 &&  2.13 &  2.15 &&  1.16 &  1.17 \\
20.5 &&  2.55 &  2.57 &&  2.00 &  2.02 &&  2.34 &  2.36 &&  1.41 &  1.42 \\
21.0 &&  2.74 &  2.76 &&  2.18 &  2.21 &&  2.53 &  2.56 &&  1.65 &  1.67 \\
21.5 &&  2.92 &  2.95 &&  2.35 &  2.38 &&  2.71 &  2.74 &&  1.89 &  1.90 \\
22.0 &&  3.09 &  3.13 &&  2.51 &  2.55 &&  2.88 &  2.92 &&  2.12 &  2.14 \\
22.5 &&  3.25 &  3.29 &&  2.65 &  2.70 &&  3.04 &  3.09 &&  2.35 &  2.37 \\
23.0 &&  3.40 &  3.45 &&  2.79 &  2.85 &&  3.19 &  3.24 &&  2.58 &  2.60 \\
23.5 &&  3.55 &  3.61 &&  2.91 &  2.98 &&  3.34 &  3.40 &&  2.80 &  2.82 \\
24.0 &&  3.70 &  3.76 &&  3.02 &  3.10 &&  3.49 &  3.55 &&  3.02 &  3.05 \\
24.5 &&  3.87 &  3.92 &&  3.13 &  3.22 &&  3.66 &  3.71 &&  3.23 &  3.27 \\
25.0 &&  4.04 &  4.09 &&  3.24 &  3.33 &&  3.83 &  3.88 &&  3.43 &  3.48 \\
25.5 &&  4.22 &  4.27 &&  3.35 &  3.44 &&  4.01 &  4.06 &&  3.62 &  3.69 \\
26.0 &&  4.39 &  4.46 &&  3.46 &  3.56 &&  4.18 &  4.25 &&  3.80 &  3.89 \\
26.5 &&  4.55 &  4.65 &&  3.60 &  3.69 &&  4.34 &  4.44 &&  3.97 &  4.08 \\
27.0 &&  4.67 &  4.83 &&  3.74 &  3.83 &&  4.46 &  4.62 &&  4.14 &  4.26 \\
27.5 &&  4.76 &  4.99 &&  3.89 &  3.98 &&  4.55 &  4.78 &&  4.30 &  4.43 \\
28.0 &&  4.83 &  5.12 &&  4.03 &  4.15 &&  4.62 &  4.91 &&  4.46 &  4.60 \\
28.5 &&  4.89 &  5.22 &&  4.13 &  4.32 &&  4.67 &  5.01 &&  4.61 &  4.75 \\
29.0 &&  4.93 &  5.30 &&  4.21 &  4.47 &&  4.71 &  5.09 &&  4.75 &  4.91 \\
\hline
\end{tabular}
\end{center}
\clearpage

\begin{center}
{\bf TABLE 4}

$I$-band No-evolution Models with a Local Hole \\
(log counts deg$^{-2}$ mag$^{-1}$)
\vskip 0.1in
\small
\begin{tabular}{ccccccccccccc}
\hline
\hline
   $I_{814}$ &&\multicolumn{2}{c}{All}&&\multicolumn{2}{c}{E/S0}&&\multicolumn{2}{c}{Spiral}&&\multicolumn{2}{c}{Irr/Pec} \\
\hline
10.5 && -1.72 & -1.72 && -1.98 & -1.98 && -1.92 & -1.92 && -3.43 & -3.43 \\
11.0 && -1.42 & -1.42 && -1.69 & -1.69 && -1.63 & -1.63 && -3.13 & -3.13 \\
11.5 && -1.13 & -1.13 && -1.39 & -1.39 && -1.33 & -1.33 && -2.84 & -2.84 \\
12.0 && -0.84 & -0.84 && -1.10 & -1.10 && -1.04 & -1.04 && -2.54 & -2.54 \\
12.5 && -0.55 & -0.55 && -0.81 & -0.81 && -0.75 & -0.75 && -2.24 & -2.24 \\
13.0 && -0.26 & -0.26 && -0.53 & -0.53 && -0.46 & -0.46 && -1.95 & -1.95 \\
13.5 &&  0.03 &  0.03 && -0.25 & -0.24 && -0.18 & -0.18 && -1.65 & -1.65 \\
14.0 &&  0.31 &  0.31 &&  0.04 &  0.04 &&  0.11 &  0.11 && -1.36 & -1.36 \\
14.5 &&  0.60 &  0.60 &&  0.34 &  0.34 &&  0.39 &  0.39 && -1.07 & -1.07 \\
15.0 &&  0.89 &  0.89 &&  0.65 &  0.65 &&  0.69 &  0.69 && -0.77 & -0.77 \\
15.5 &&  1.21 &  1.21 &&  0.97 &  0.97 &&  1.00 &  1.00 && -0.46 & -0.46 \\
16.0 &&  1.52 &  1.52 &&  1.28 &  1.28 &&  1.32 &  1.32 && -0.15 & -0.15 \\
16.5 &&  1.83 &  1.83 &&  1.57 &  1.57 &&  1.62 &  1.62 &&  0.17 &  0.17 \\
17.0 &&  2.11 &  2.12 &&  1.83 &  1.84 &&  1.90 &  1.91 &&  0.49 &  0.49 \\
17.5 &&  2.37 &  2.38 &&  2.07 &  2.08 &&  2.16 &  2.17 &&  0.79 &  0.79 \\
18.0 &&  2.61 &  2.62 &&  2.29 &  2.31 &&  2.40 &  2.42 &&  1.08 &  1.08 \\
18.5 &&  2.83 &  2.85 &&  2.50 &  2.53 &&  2.63 &  2.65 &&  1.36 &  1.37 \\
19.0 &&  3.04 &  3.07 &&  2.69 &  2.72 &&  2.84 &  2.86 &&  1.63 &  1.63 \\
19.5 &&  3.23 &  3.27 &&  2.86 &  2.91 &&  3.03 &  3.06 &&  1.88 &  1.89 \\
20.0 &&  3.41 &  3.46 &&  3.02 &  3.08 &&  3.21 &  3.25 &&  2.12 &  2.14 \\
20.5 &&  3.58 &  3.63 &&  3.16 &  3.24 &&  3.37 &  3.43 &&  2.36 &  2.38 \\
21.0 &&  3.72 &  3.80 &&  3.29 &  3.38 &&  3.52 &  3.59 &&  2.58 &  2.61 \\
21.5 &&  3.86 &  3.94 &&  3.40 &  3.51 &&  3.65 &  3.73 &&  2.78 &  2.83 \\
22.0 &&  3.98 &  4.08 &&  3.50 &  3.62 &&  3.77 &  3.87 &&  2.98 &  3.03 \\
22.5 &&  4.09 &  4.20 &&  3.59 &  3.73 &&  3.87 &  3.99 &&  3.18 &  3.23 \\
23.0 &&  4.18 &  4.32 &&  3.67 &  3.82 &&  3.97 &  4.10 &&  3.36 &  3.42 \\
23.5 &&  4.28 &  4.42 &&  3.75 &  3.92 &&  4.07 &  4.21 &&  3.55 &  3.61 \\
24.0 &&  4.37 &  4.52 &&  3.82 &  4.00 &&  4.15 &  4.31 &&  3.73 &  3.80 \\
24.5 &&  4.45 &  4.62 &&  3.88 &  4.09 &&  4.24 &  4.41 &&  3.90 &  3.98 \\
25.0 &&  4.53 &  4.72 &&  3.94 &  4.16 &&  4.32 &  4.50 &&  4.07 &  4.16 \\
25.5 &&  4.62 &  4.81 &&  4.00 &  4.24 &&  4.40 &  4.59 &&  4.24 &  4.34 \\
26.0 &&  4.70 &  4.90 &&  4.06 &  4.31 &&  4.48 &  4.68 &&  4.41 &  4.51 \\
26.5 &&  4.78 &  5.00 &&  4.11 &  4.38 &&  4.56 &  4.78 &&  4.57 &  4.68 \\
27.0 &&  4.84 &  5.10 &&  4.17 &  4.44 &&  4.63 &  4.88 &&  4.73 &  4.84 \\
27.5 &&  4.90 &  5.19 &&  4.22 &  4.51 &&  4.68 &  4.97 &&  4.88 &  5.00 \\
28.0 &&  4.95 &  5.27 &&  4.26 &  4.58 &&  4.73 &  5.06 &&  5.03 &  5.15 \\
28.5 &&  4.99 &  5.34 &&  4.29 &  4.65 &&  4.76 &  5.12 &&  5.17 &  5.30 \\
29.0 &&  5.02 &  5.39 &&  4.31 &  4.71 &&  4.80 &  5.18 &&  5.31 &  5.45 \\
\hline
\end{tabular}
\end{center}

\clearpage

\clearpage

\figcaption[]
{A sample of images representing the range of difficulty encountered in galaxy
classification.  The middle column shows a galaxy of average difficulty
in each class, while the left and right columns show approximately the
least difficult tenth and the most challenging tenth of the sample, respectively.
\label{fig1}}

\figcaption[]
{The dependence of the SSRS2 luminosity function 
on galaxy morphology.  The dashed line connects the 
SWML estimates for the sample as a whole.
\label{fig2}}

\figcaption[]
{Confidence intervals (1$\sigma$) on the Schechter shape parameters for
different types in the SSRS2.
\label{fig3}}

\figcaption[]
{Redshift maps of the SSRS2.  The left column includes the entire range
of redshifts covered by the survey: $0 \le cz \le 20,000 \kms$.  The
upper panel shows the southern half of the survey; the lower panel
shows the northern half.  The panels on the right are enlarged versions
of the left-hand panels covering only the redshift range $0 \le cz \le
3,000 \kms$.
\label{fig4}}

\figcaption[]
{Luminosity functions for each morphology computed using different
assumptions about the local galaxy velocity field.  The histogram
reproduces the original luminosity functions from Figure 2.  The 
open squares represent case 1 (see text); filled circles represent
case 2.
\label{fig5}}

\figcaption[]
{A comparison of recent measurements of the LF divided by morphological type.
Solid lines and open squares represent the SSRS2; dashed lines represent the CfA Survey
and dotted lines represent the Stromlo-APM.  
\label{fig6}}

\figcaption[]
{A representative compilation of galaxy counts in the $B$ band covering
the observed range of apparent magnitudes.  Solid triangles indicate
the SSRS2, open circles the APM (Maddox \etal 1990), triangles the CCD counts of Metcalfe
\etal (1991), and open squares the Hubble and ASU Deep Fields (Williams \etal 1996, Odewahn \etal 1996).  
In each case, magnitudes have been transformed
approximately to $b_j$, which roughly matches the SSRS2 system. The solid line is the fiducial
no-evolution model for the galaxy counts based on the shape of the 
SSRS2 LF but including a normalization which increases by a factor
of 2 at $z \sim 0.1$.
\label{fig7}}

\figcaption[]
{A comparison of the expected redshift distributions to the measured redshift
distributions for subsamples of the SSRS2.  From top to bottom, the panels
represent the full SSRS2, SSRS2 South, and SSRS2 North.  The histograms in the
left-hand panels are the observed redshift distributions.  The solid lines
in these panels are the expectations based on a uniform distribution of 
galaxies and the overall luminosity function of the SSRS2.  The predicted
distributions differ only by a scale factor corresponding to 
the solid angle covered by each subsample; the luminosity function is the
same in each panel.  The points in the right-hand
panels are the ratio of the observed redshift distribution to the predicted
one.  
\label{fig8}}

\figcaption[]
{No-evolution predictions for the Hubble Deep Field galaxy counts
from Abraham \etal (1996).  Solid lines are for $q_0=0.05$; dashed
lines indicate $q_0=0.5$.  For the irregulars, two pairs of lines
are shown: the lower pair is computed using a cutoff in the LF at $M_B=-14$.
For the upper pair, we extrapolate the Schechter function to $M_B=-10$.  
\label{fig9}}


\clearpage

\plotone{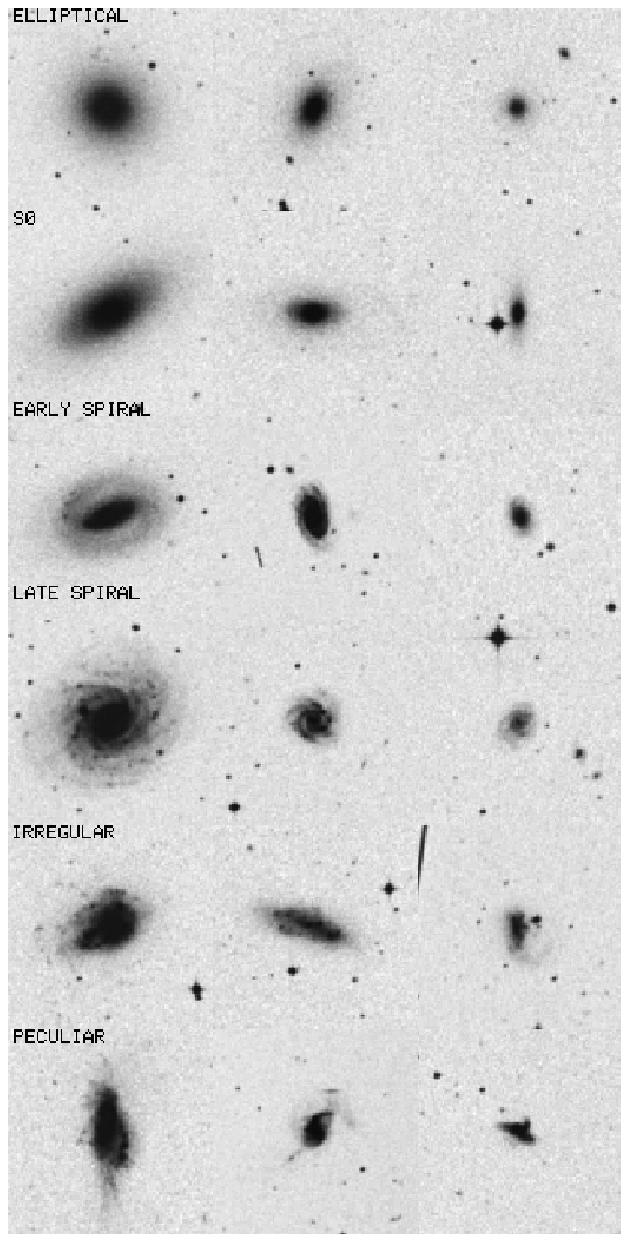}
\hfill Figure 1 \hfill

\clearpage

\plotone{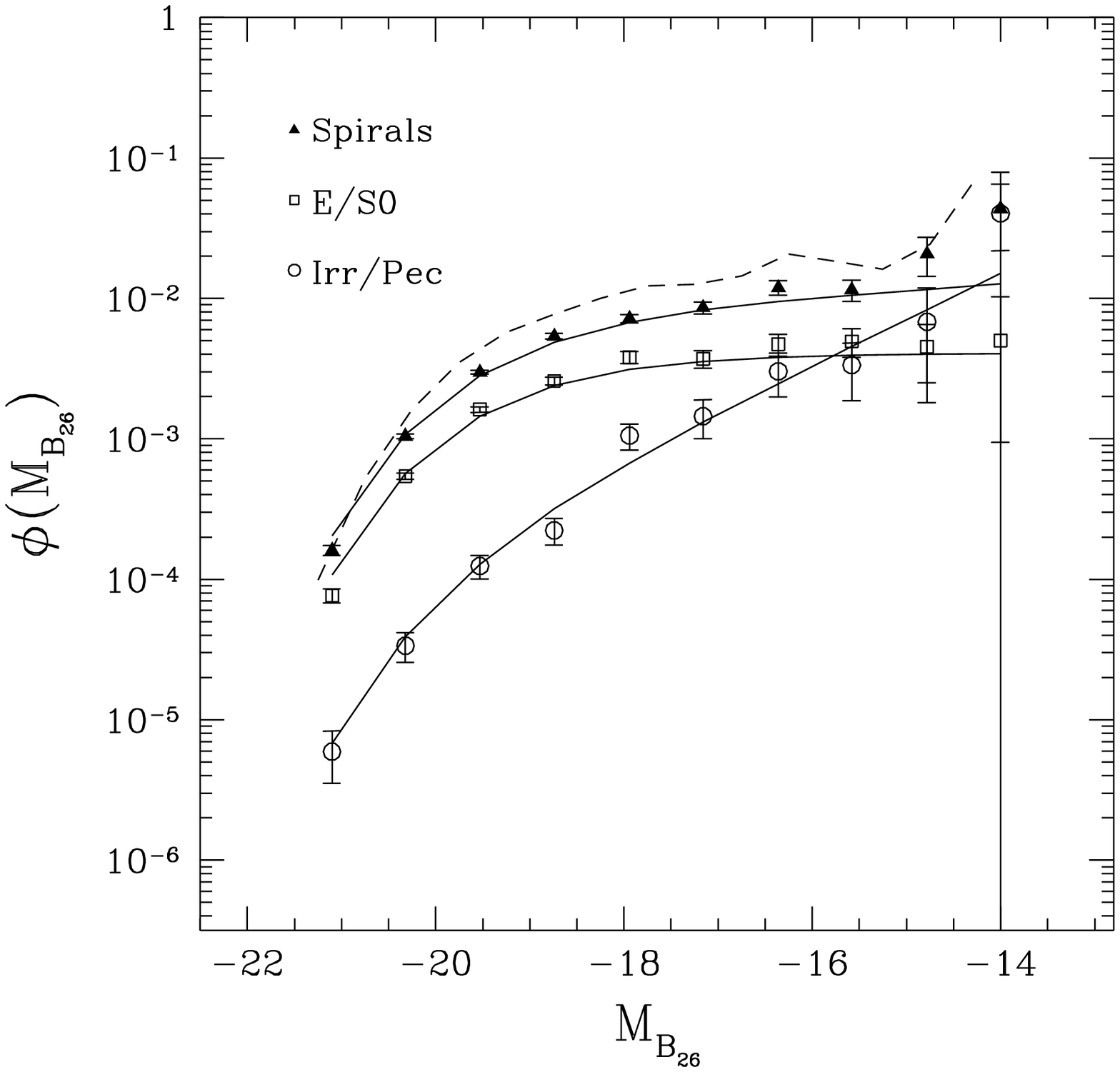}
\hfill Figure 2 \hfill

\clearpage

\plotone{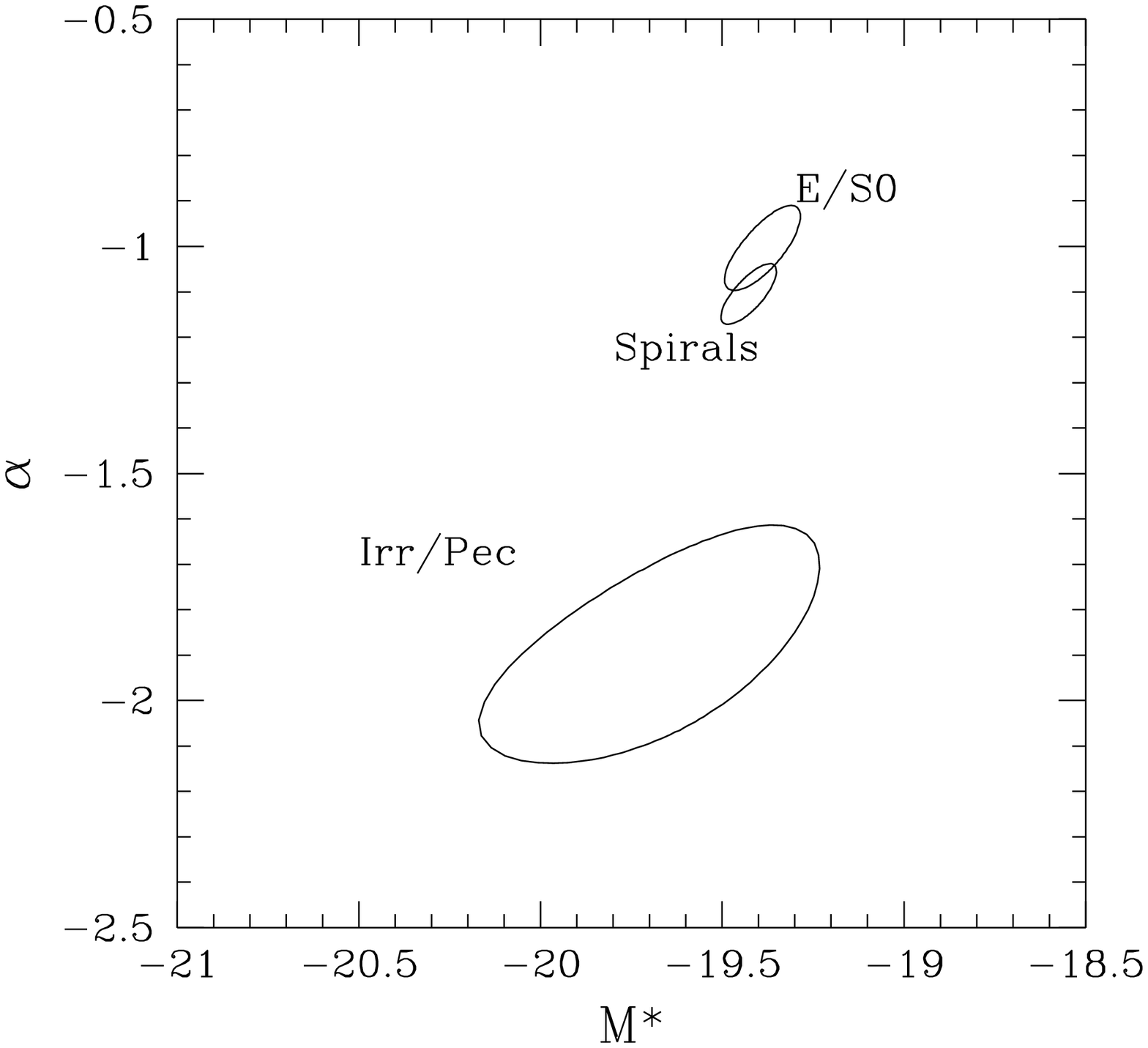}
\hfill Figure 3 \hfill

\clearpage

\plotone{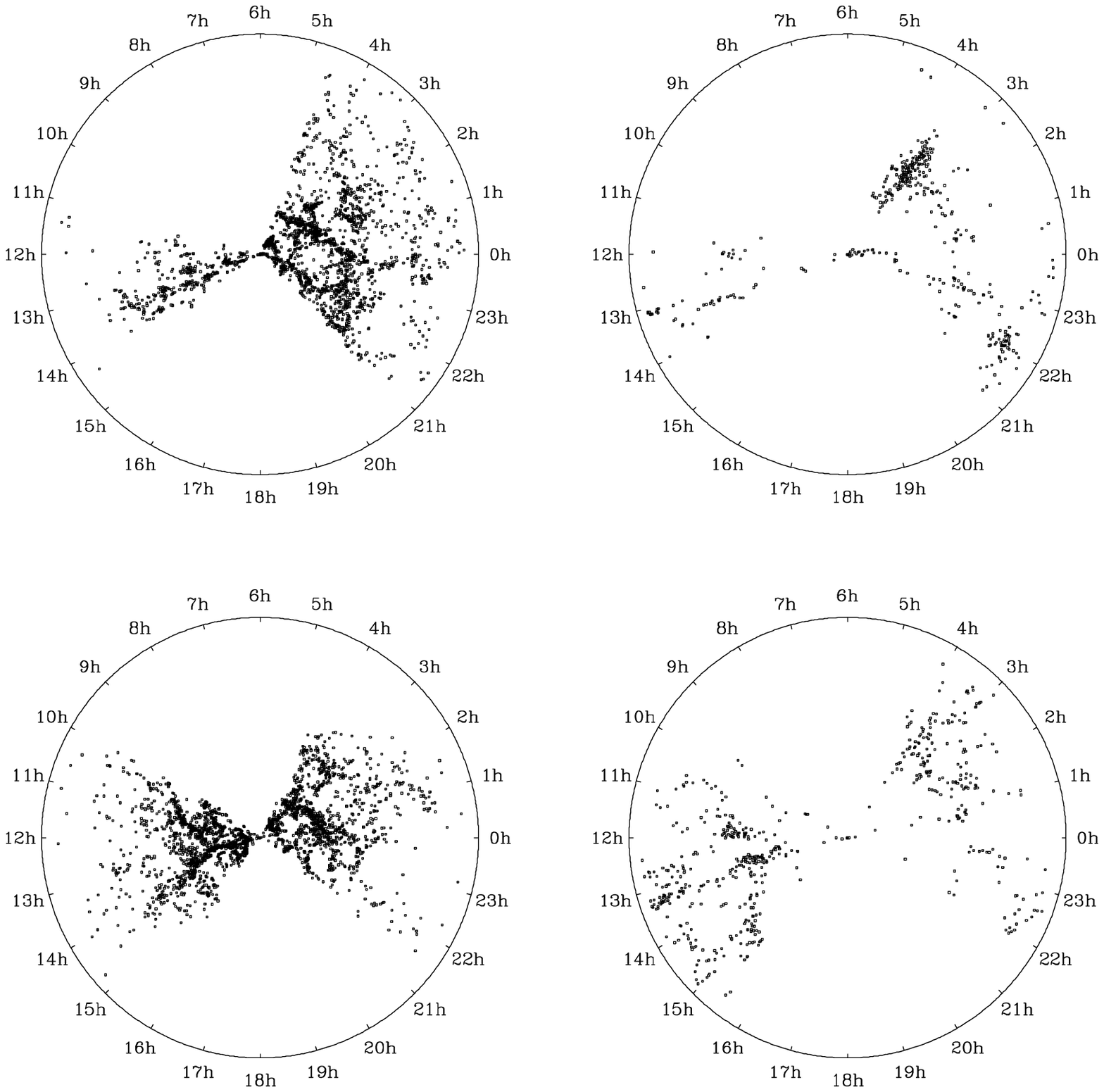}
\hfill Figure 4 \hfill

\clearpage

\plotone{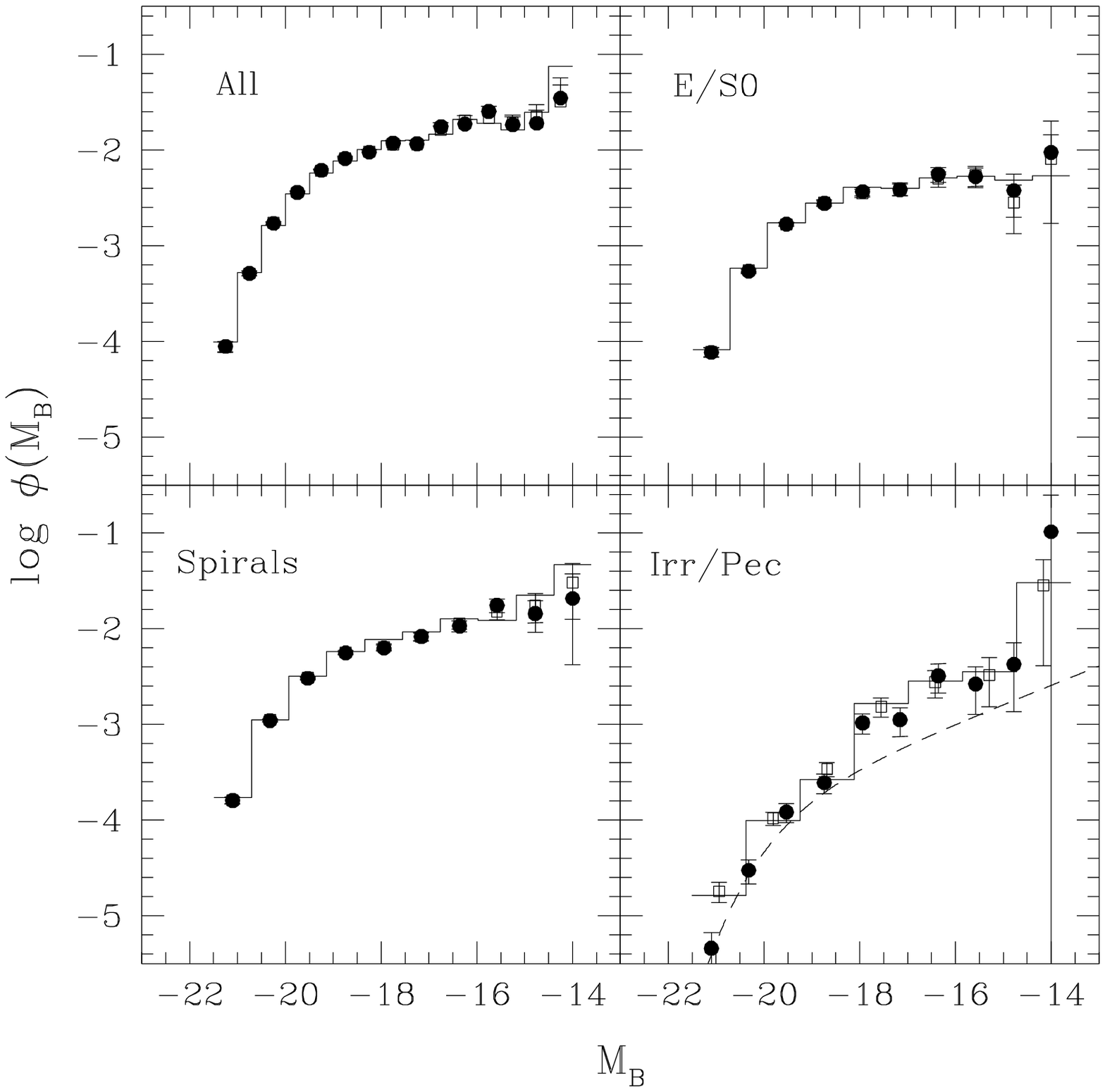}
\hfill Figure 5 \hfill

\clearpage

\plotone{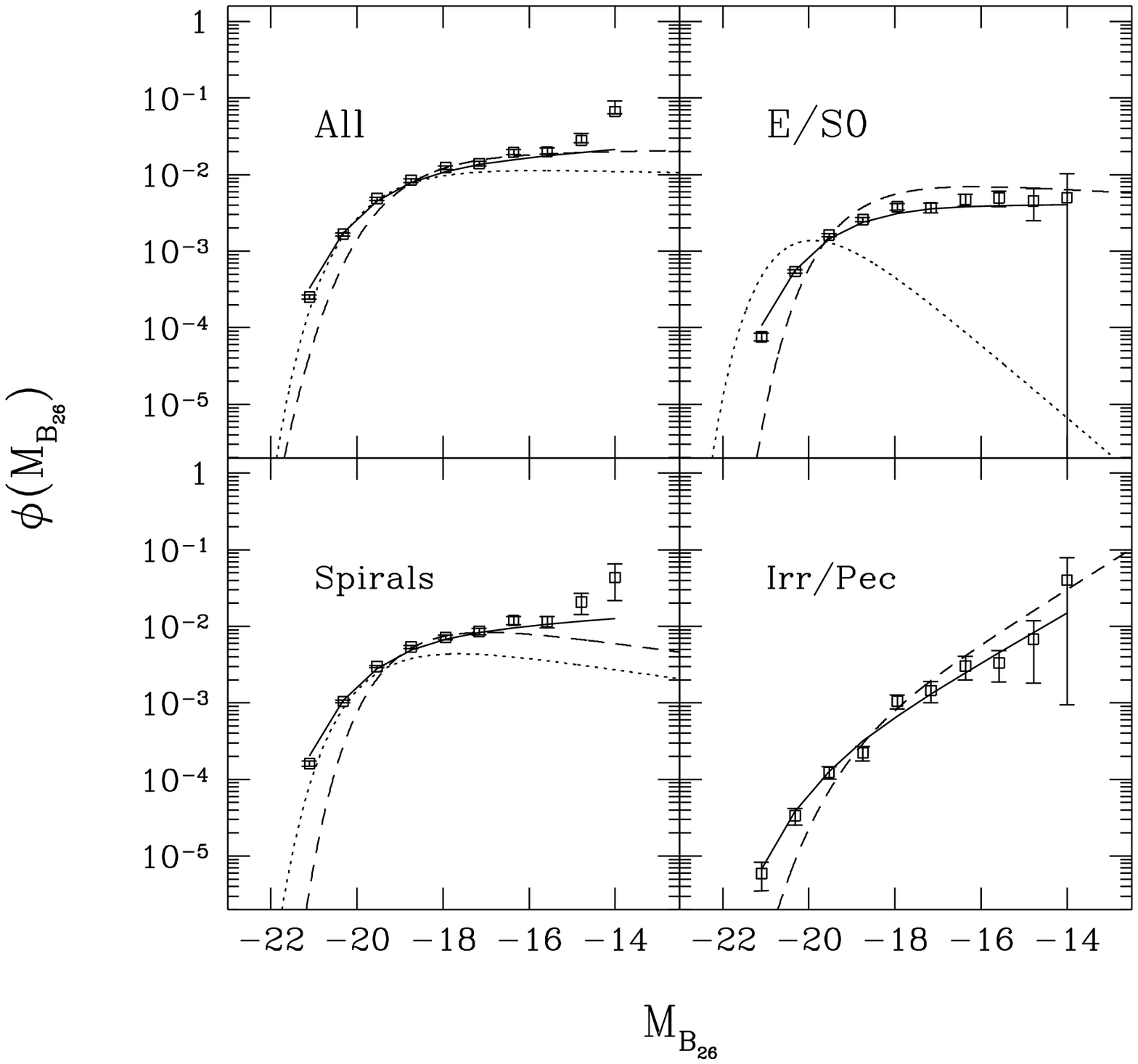}
\hfill Figure 6 \hfill

\clearpage

\plotone{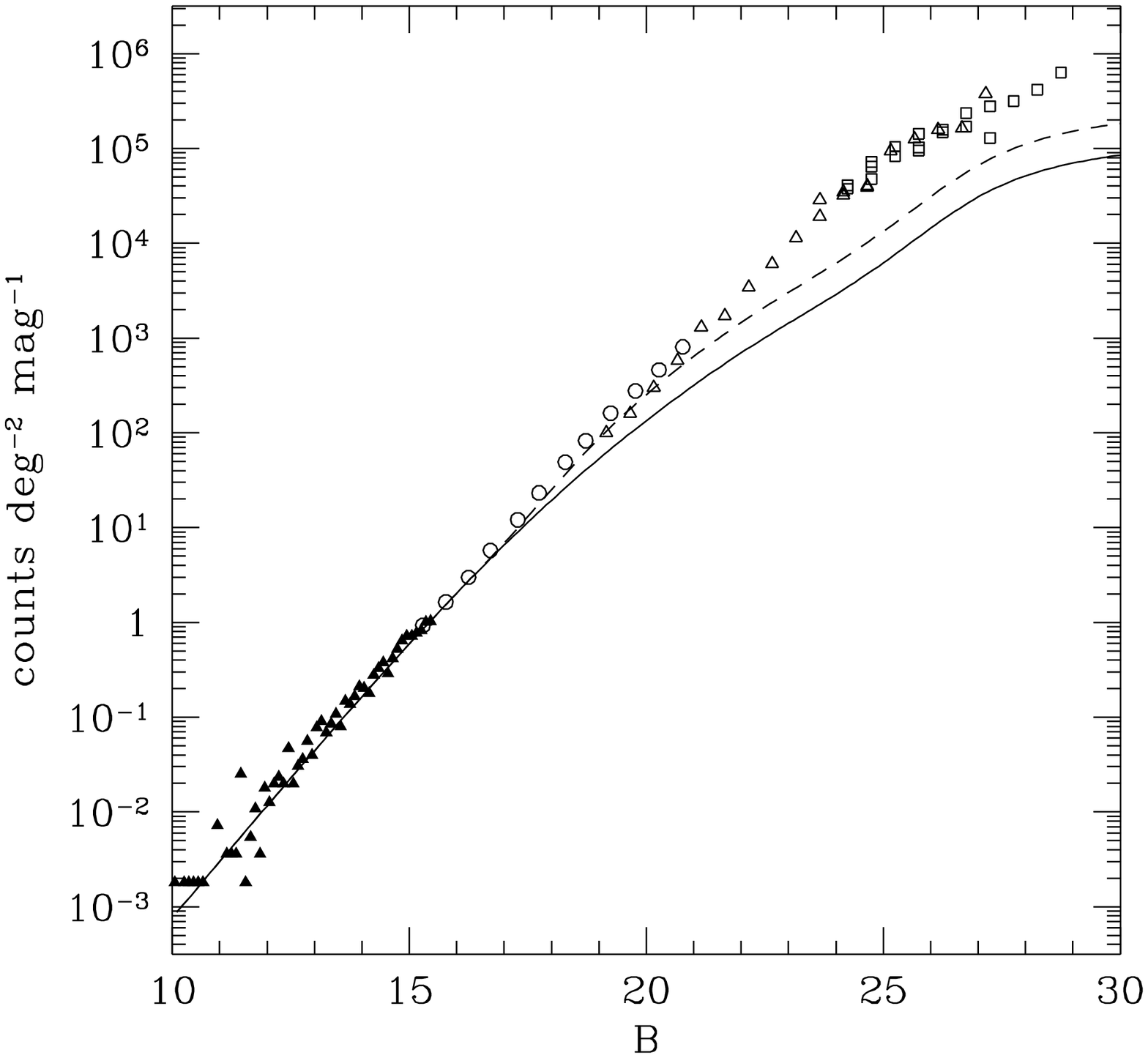}
\hfill Figure 7 \hfill

\clearpage

\plotone{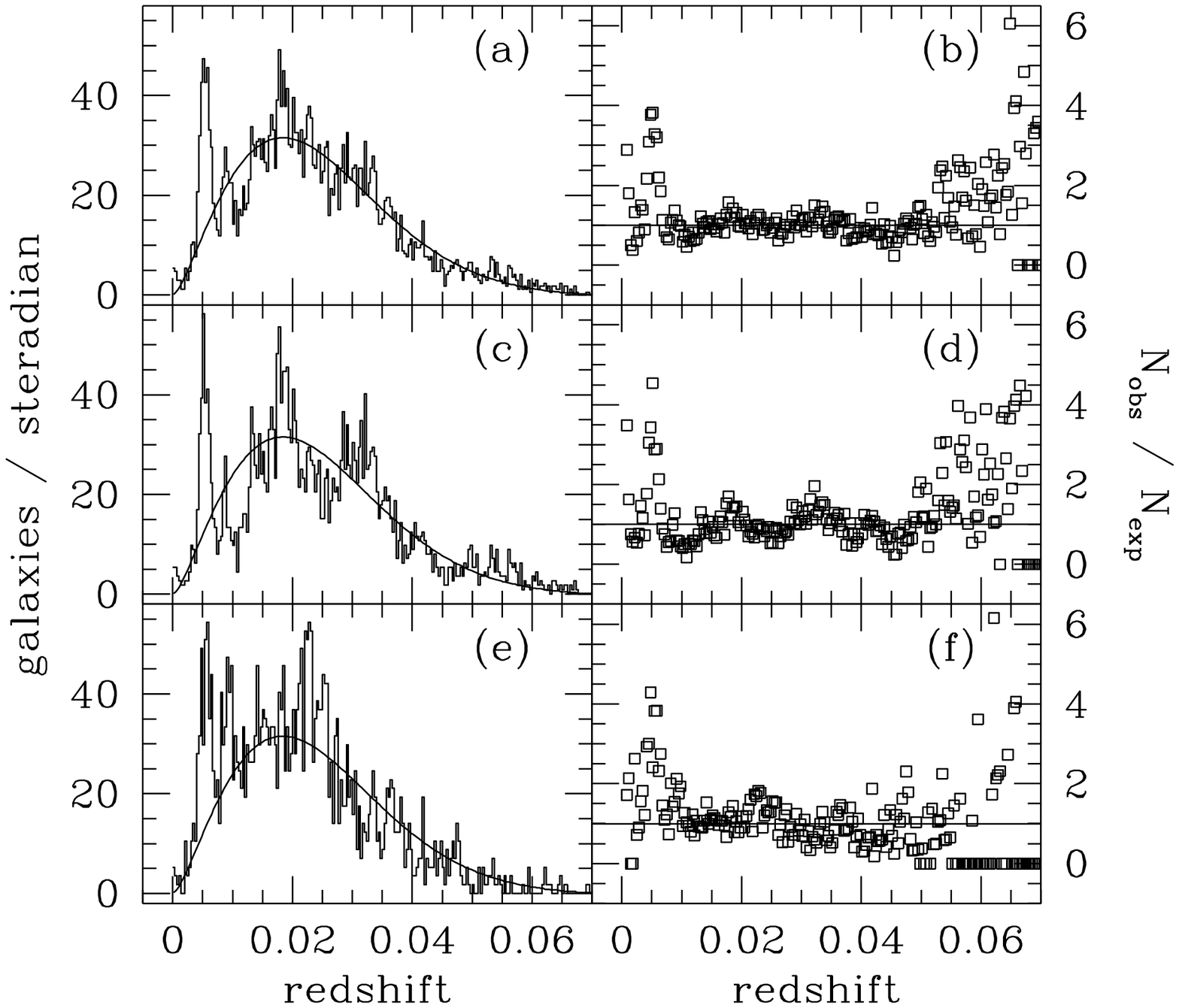}
\hfill Figure 8 \hfill

\clearpage

\plotone{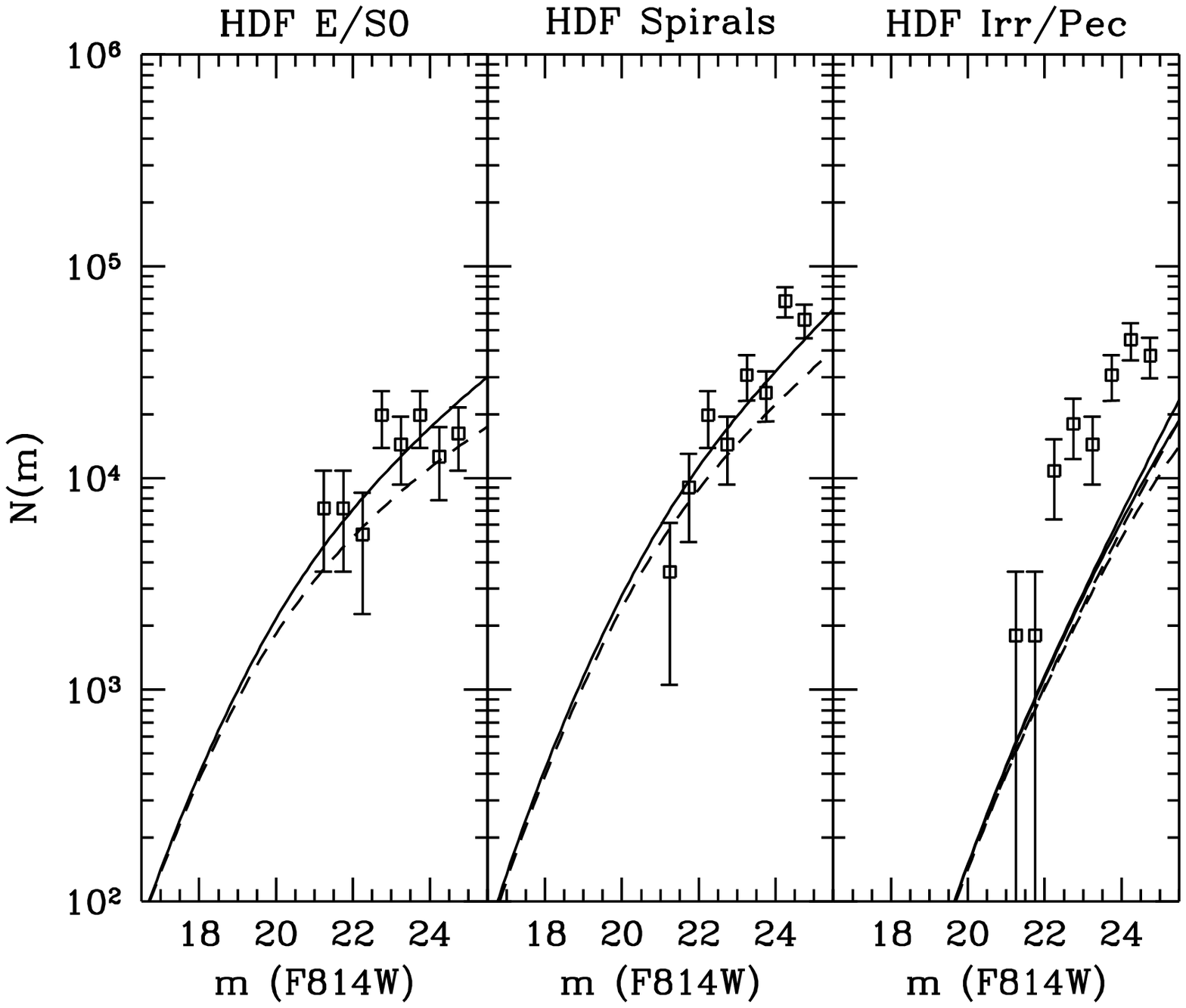}
\hfill Figure 9 \hfill

\clearpage

\end{document}